\documentclass{article}
\usepackage{amsmath,amssymb,amsfonts}
\usepackage{graphicx}

\def\rank{\mathrm{rank}}

\def\const{\mathrm{const}}

\newtheorem{theorem}{Theorem}
\newtheorem{proposition}{Proposition}
\newtheorem{corollary}{Corollary}
\newtheorem{lemma}{Lemma}

\newtheorem{example}{Example}
\newtheorem{definition}{Definition}

\newenvironment{proof}[1][Proof]{\noindent\textit{#1.} }{\hfill$\Box$\medskip}

\title{Geometry of Integrable Billiards and Pencils of
Quadrics}

\author{Vladimir Dragovi\'c and Milena Radnovi\'c
\footnote{on leave at The Weizmann Institute of
Science, Rehovot, Israel}
\\{\small Mathematical Institute SANU,
Belgrade, Serbia and Montenegro}\\{\small{\tt
vladad@mi.sanu.ac.yu, milena@mi.sanu.ac.yu}}}

\date{}

\begin{document}

\maketitle

\centerline{\large\it to be published in:}

\centerline{\large \it Journal de Math\'ematiques Pures et
Appliqu\'ees}

\

\

\

\begin{abstract} We study the deep interplay between geometry of
quadrics in $d$-di\-men\-sional space and the dynamics of related
integrable billiard systems. Various generalizations of Poncelet
theorem are reviewed. The corresponding analytic conditions of
Cayley's type are derived giving the full description of
periodical billiard trajectories; among other cases, we consider
billiards in arbitrary dimension $d$ with the boundary consisting
of arbitrary number $k$ of confocal quadrics. Several important
examples are presented in full details demonstrating the
effectiveness of the obtained results. We give a thorough analysis
of classical ideas and results of Darboux and methodology of
Lebesgue, and prove their natural generalizations, obtaining new
interesting properties of pencils of quadrics. At the same time,
we show essential connections between these classical ideas and
the modern algebro-geometric approach in the integrable systems
theory.
\end{abstract}

\newpage

\tableofcontents

\newpage

\section{Introduction}

In his {\it Trait\'{e} des propri\'et\'es projectives des figures}
\cite{Pon}, Poncelet proved one of most beautiful and most
important claims of the 19th century geometry. Suppose that two
ellipses are given in the plane, together with a closed polygonal
line inscribed in one of them and circumscribed about the other
one. Then, Poncelet theorem states that infinitely many such
closed polygonal lines exist –-- every point of the first ellipse
is a vertex of such a polygon. Besides, all these polygons have
the same number of sides. Poncelet's proof was purely geometrical,
synthetic. Later, using the addition theorem for elliptic
functions, Jacobi gave another proof of the theorem \cite{Jac}.
Essentially, Poncelet theorem is equivalent to the addition
theorem and Poncelet's proof represents a synthetic way of
deriving the group structure on an elliptic curve. Another proof,
in a modern, algebro-geometrical manner, can be found in
Griffiths' and Harris' paper \cite{GH1}. There, they also gave an
interesting generalization of the Poncelet theorem to the
three-dimensional case, considering polyhedral surfaces both
inscribed and circumscribed about two quadrics.

\smallskip

A natural question connected with Poncelet theorem is to find an
analytical condition determining, for two given conics, if an
$n$-polygon inscribed in one and circumscribed about the second
conic exists. In a short paper \cite{Cay}, Cayley derived such a
condition, using the theory of Abelian integrals. Inspired by this
paper, Lebesgue translated Cayley's proof to the language of
geometry. Lebesgue's proof of Cayley's condition, derived by
methods of projective geometry and algebra, can be found in his
book {\it Les coniques} \cite{Leb}. Griffiths and Harris derived
Cayley theorem by finding an analytical condition for points of
finite order on an elliptic curve \cite{GH2}.

\smallskip

It is worth emphasizing that Poncelet, in fact, proved a statement
that is much more general than the famous Poncelet theorem
\cite{Ber, Pon}, then deriving the latter as a corollary. Namely,
he considered $n+1$ conics of a pencil in the projective plane. If
there exists an $n$-polygon with vertices lying on the first of
these conics and each side touching one of the other $n$ conics,
then infinitely many such polygons exist. We shall refer to this
statement as {\it Complete Poncelet theorem} (CPT) and call such
polygons {\it Poncelet polygons}. We are going to follow here
mostly the presentation of Lebesgue from \cite{Leb}, which is, as
we learned from M.\ Berger, quite close to one of two Poncelet's
original proofs.

\smallskip

A nice historical overview of the Poncelet theorem, together with
modern proofs and remarks is given in \cite{BKOR}. Various
classical theorems of Poncelet type with short modern proofs
reviewed in \cite{BB}, while the algebro-geometrical approach to
families of Poncelet polygons via modular curves is given in
\cite{BM, Jak}.

\smallskip

Poncelet theorem has a nice mechanical interpretation. {\it
Elliptical billiard} \cite{KT} is a dynamical system where a
material point of the unit mass is moving with a constant velocity
inside an ellipse and obeying the reflection law at the boundary,
i.e.\ having congruent impact and reflection angles with the
tangent line to the ellipse at any bouncing point. It is also
assumed that the reflection is absolutely elastic. It is well
known that any segment of a given elliptical billiard trajectory
is tangent to the same conic, confocal with the boundary
\cite{CCS}. If a trajectory becomes closed after $n$ reflections,
then Poncelet theorem implies that any trajectory of the billiard
system, which shares the same caustic curve, is also periodic with
the period $n$.

\smallskip

Complete Poncelet theorem also has a mechanical meaning. The
configuration dual to a pencil of conics in the plane is a family
of confocal second order curves \cite{Ar}. Let us consider the
following, a little bit unusual billiard. Suppose $n$ confocal
conics are given. A particle is bouncing on each of these $n$
conics respectively. Any segment of such a trajectory is tangent
to the same conic confocal with the given $n$ curves. If the
motion becomes closed after $n$ reflections, then, by Complete
Poncelet theorem, any such a trajectory with the same caustic is
also closed.

\smallskip

The statement dual to Complete Poncelet theorem can be generalized
to the $d$-dimensional space \cite{CCS}. Suppose vertices of the
polygon $x_1x_2\dots x_n$ are respectively placed on confocal
quadric hyper-surfaces $\mathcal Q_1$, $\mathcal Q_2$, \dots,
$\mathcal Q_n$ in the $d$-dimensional Eucledean space, with
consecutive sides obeying the reflection law at the corresponding
hyper-surface. Then all sides are tangent to some quadrics
$\mathcal Q^1$, \dots, $\mathcal Q^{d-1}$ confocal with
$\{\mathcal Q_i\}$; for the hyper-surfaces $\{\mathcal
Q_i,\mathcal Q^j\}$, an infinite family of polygons with the same
properties exist.

\smallskip

But, more than one century before these quite recent results,
Darboux proved the generalization of Poncelet theorem for a
billiard within an ellipsoid in the three-dimensional space
\cite{Dar1}. It seems that his work on this topic is completely
forgot nowadays.

\smallskip

It is natural to search for a Cayley-type condition related to
some of generalizations of Poncelet theorem. The authors derived
such conditions for the billiard system inside an ellipsoid in the
Eucledean space of arbitrary finite dimension \cite{DR1, DR2}. In
our recent note \cite{DR3}, algebro-geometric conditions for
existence of periodical billiard trajectories within $k$ quadrics
in $d$-dimensional Euclidean space were announced. The aim of the
present paper is to give full explanations and proofs of these
results together with several important examples and improvements.
The second important goal of this paper is to offer a thorough
historical overview of the subject with a special attention on the
detailed analysis of ideas and contributions of Darboux and
Lebesgue. While Lebesgue's work on this subject has been, although
rarely, mentioned by experts, on the other hand, it seems to us
that relevant Darboux's ideas are practically unknown in the
contemporary mathematics. We give natural higher dimensional
generalizations of the ideas and results of Darboux and Lebesgue,
providing the proofs also in the low-dimensional cases if they
were omitted in the original works. Beside other results,
interesting new properties of pencils of quadrics are established
-- see Theorems  \ref{th:virt.refl} and \ref{uopsten.lebeg}. The
latter gives a nontrivial generalization of the Basic Lemma.

\smallskip

This paper is organized as follows. In the next section, a short
review of Lebesgue's results from \cite{Leb} is given, followed by
their application to the case of the billiard system between two
confocal ellipses. Section \ref{hyper} contains algebro-geometric
discussions which will be applied in the rest of the paper. In
Section \ref{k.quad}, we give analytic conditions for periodicity
of billiard motion inside a domain bounded by several confocal
quadrics in the Euclidean space of arbitrary dimension. The
complexity of the problem of billiard motion within several
quadrics is well known, even in the real case, and it is induced
by multivaluedness of the billiard mapping. Thus, to establish a
correct setting of the problem, we introduce basic notions of
reflections {\it from inside} and {\it from outside} a quadric
hyper-surface, and we define {\it the billiard ordered game}. The
corresponding closeness conditions are derived, together with
examples and discussions. In Section \ref{discrete}, we consider
the elliptical billiard as a discrete-time dynamical system, and,
applying the Veselov's and Moser's algebro-geometric integration
procedure, we derive the periodicity conditions. The obtained
results are compared with those from Section \ref{k.quad}. In
Section \ref{on.quad}, we give an algebro-geometric description of
periodical trajectories of the billiard motion on quadric
hyper-surfaces, we study the behaviour of geodesic lines after the
reflection at a confocal quadric and derive a new porism of
Poncelet type. In Section \ref{virtual}, we define the virtual
reflection configuration, prove Darboux's statement on virtual
billiard trajectories, generalize it to arbitrary dimension and
study related geometric questions. In Section \ref{gen.lebeg}, we
formulate and prove highly nontrivial generalization of the Basic
Lemma (Lemma \ref{lebeg.lema}), giving a new important geometric
property of dual pencils of quadrics. In that section, we also
introduce and study the {\it generalized Cayley curve}, a natural
higher-dimensional generalization of {\it the Cayley cubic}
studied by Lebesgue. In this way, in Section \ref{gen.lebeg} the
most important tools of Lebesgue's study are generalized. Further
development of this line will be presented in separate publication
\cite{DR4}. In Appendix 1, we review some known classes of
integrable potential perturbations of elliptical billiards,
emphasizing connections with Appell hypergeometric functions and
Liouville surfaces. Finally, in Appendix 2, we present the related
Darboux's results considering a generalization of Poncelet theorem
to Liouville surfaces, giving a good basis for a study of the
geometry of periodic trajectories appearing in the perturbed
systems from Appendix 1.

\section{Planar Case: $d=2$, $k$ -- Arbitrary}\label{planar}

First of all, we consider the billiard system within $k$ confocal
ellipses in the 2-dimensional plane. In such a system, the
billiard particle bounces sequentially of these confocal ellipses.
We wish to get the analytical description of periodical
trajectories of such a system.

\smallskip

Following Lebesgue, let us consider polygons inscribed in a conic
$\Gamma$, whose sides are tangent to $\Gamma_1,\dots,\Gamma_k$,
where $\Gamma, \Gamma_1,\dots,\Gamma_k$ all belong to a pencil of
conics. In the dual plane, such polygons correspond to billiard
trajectories having caustic $\Gamma^*$ with bounces on
$\Gamma_1^*,\dots,\Gamma_k^*$. The main object of Lebesgue's
analysis is the cubic Cayley curve, which parametrizes contact
points of tangents drawn from a given point to all conics of the
pencil.

\subsection{Full Poncelet Theorem}

\noindent{\bf Basic Lemma.} Next lemma is the main step in the
proof of full Poncelet theorem. If one Poncelet polygon is given,
this lemma enables us to construct every Poncelet polygon with
given initial conditions. Also, the lemma is used in deriving of a
geometric condition for the existence of a Poncelet polygon.

\begin{lemma} {\rm \cite{Leb}}\label{lebeg.lema}
Let $\mathcal F$ be a pencil of conics in
the projective plane and $\Gamma$ a conic from this pencil. Then
there exist quadrangles whose vertices $A$, $B$, $C$, $D$ are on
$\Gamma$ such that three pairs of its non-adjacent sides $AB$,
$CD$; $AC$, $BD$; $AD$, $BC$ are tangent to three conics of
$\mathcal F$. Moreover, the six contact points all lie on a line
$\Delta$. Any such a quadrangle is determined by two sides and the
corresponding contact points.
\end{lemma}

Let $\Gamma$, $\Gamma_1$, $\Gamma_2$, $\Gamma_3$ be conics of a
pencil and $ABC$ a Poncelet triangle corresponding to these
conics, such that its vertices lie on $\Gamma$ and sides $AB$,
$BC$, $CA$ touch $\Gamma_1$, $\Gamma_2$, $\Gamma_3$ respectively.
This lemma gives us a possibility to construct triangle $ABD$
inscribed in $\Gamma$ whose sides $AB$, $BD$, $DA$ touch conics
$\Gamma_1$, $\Gamma_3$, $\Gamma_2$ respectively. In a similar
fashion, for a given Poncelet polygon, we can, applying Lemma 1,
construct another polygon which corresponds to the same conics,
but its sides are tangent to them in different order.

\

\noindent{\bf Circumscribed and Tangent Polygons.} Let a triangle
$ABC$ be inscribed in a conic $\Gamma$ and sides $BC$, $AC$, $AB$
touch conics $\Gamma_1$, $\Gamma_2$, $\Gamma_3$ of the pencil
$\mathcal F$ at points $M$, $N$, $P$ respectively. According to
the Lemma, there are two possible cases: either points $M$, $N$,
$P$ are collinear, when we will say that the triangle is {\sl
tangent} to $\Gamma_1$, $\Gamma_2$, $\Gamma_3$; or the line $MN$
is intersecting $AB$ at a point $S$ which is a harmonic conjugate
to $P$ with respect to the pair $A$, $B$, then we say that the
triangle is {\sl circumscribed} about $\Gamma_1$, $\Gamma_2$,
$\Gamma_3$.

\smallskip

Let $ABCD \dots KL$ be a polygon inscribed in $\Gamma$ whose sides
touch conics $\Gamma_1$, \dots, $\Gamma_n$ of the pencil $\mathcal
F$ respectively. Denote by $(AC)$ a conic such that $\triangle
ABC$ is circumscribed about $\Gamma_1$, $\Gamma_2$, $(AC)$, by
$(AD)$ a conic such that $\triangle ACD$ is circumscribed about
$(AC)$, $\Gamma_3$, $(AD)$. Similarly, we find conics $(AE)$,
\dots, $(AK)$. The triangle $AKL$ can be tangent to the conics
$(AK)$, $\Gamma_{n-1}$, $\Gamma_n$ or circumscribed about them,
and we will say that $ABCD \dots KL$ is {\sl tangent} or,
respectively, {\sl circumscribed} about conics $\Gamma_1$, \dots,
$\Gamma_n$.

\smallskip

Further, we will be interested only in circumscribed polygons. The
Poncelet theorem does not hold for tangent triangles nor, hence,
for tangent polygons with greater number of vertices.

\begin{theorem}\label{CPT} {\rm (Complete Poncelet theorem)} Let conics
$\Gamma$, $\Gamma_1$, \dots, $\Gamma_n$ belong to a pencil
$\mathcal F$. If a polygon inscribed in $\Gamma$ and circumscribed
about $\Gamma_1$, \dots, $\Gamma_n$ exists, then infinitely many
such polygons exist.

To determine such a polygon, it is possible to give arbitrarily:

\noindent$1)$ the order which its sides touch $\Gamma_1$, \dots,
$\Gamma_n$ in; let the order be: $\Gamma_1'$, \dots, $\Gamma_n'$;

\noindent$2)$ a tangent to $\Gamma_1'$ containing one side of the
polygon;

\noindent$3)$ the intersecting point of this tangent with $\Gamma$
which will belong to the side tangent to $\Gamma_2'$.
\end{theorem}

The proof is given in \cite{Leb}.

\subsection{Cayley's Condition}

\noindent{\bf Representation of Conics of a Pencil by Points on a
Cubic Curve.} Let pencil $\mathcal F$ of conics be determined by
the curves $C=0$ and $\Gamma=0$. The equation of an arbitrary
conic of the pencil is $C+ \lambda \Gamma =0$.

\smallskip

Let $P + \lambda \Pi =0$ be the equation of the corresponding
polar lines from the point $A \in \Gamma$. The geometric place of
contact points of tangents from $A$ with conics of the pencil is
the cubic $\mathcal C: C \Pi - \Gamma P = 0$. On this cubic, any
conic of $\mathcal F$ is represented by two contact points, which
we will call {\sl representative points} of the conic. The line
determined by these two points passes through point $Z : P=0,\
\Pi=0$. There exist exactly four conics of the pencil whose
representative points coincide: the conic $\Gamma$ and three
degenerate conics with representative points $A$, $\alpha$,
$\beta$, $\gamma$. Lines $ZA$, $Z \alpha$, $Z \beta$, $Z \gamma$
are tangents to $\mathcal C$ constructed from $Z$. The tangent
line to cubic $\mathcal C$ at point $Z$ is a polar of point $A$
with respect to the conic of the pencil which contains $Z$.

\

\noindent{\bf Condition for Existence of a Poncelet Triangle.} If
triangles inscribed in $\Gamma$ and circumscribed about
$\Gamma_1$, $\Gamma_2$, $\Gamma_3$ exist, we will say that the
conics $\Gamma_1$, $\Gamma_2$, $\Gamma_3$ are {\sl joined} to
$\Gamma$. In this case, CPT states there is six such triangles
with the vertex $A \in \Gamma$. Let $ABC$ be one of them. Side
$AB$, denote it by {\it 1}, touches $\Gamma_1$ in point $m_1$.
Also, it touches another conic of the pencil, denote it by $(I)$,
in point $M$. Side $AC$, denote it by {\it 2'}, touches $\Gamma_2$
in $m_2'$. Consider the quadrangle $ABCD$ determined by $AB$, $AC$
and contact points $M$, $m_2'$. Line $Mm_2'$ meets $BC$ at
$\mu_3$, its point of tangency to $\Gamma_3$, and meets $AD$
(which we will denote by {\it 3'}) at the point $m_3'$ of tangency
to $\Gamma_3$. Triangle $ABD$ is circumscribed about $\Gamma_1$,
$\Gamma_2$, $\Gamma_3$.

\smallskip

Similarly, triangle $ACE$ can be obtained by construction of
quadrangle $ABCE$ determined by $AB$, $BC$ and contact points
$m_1$, $\mu_3$. Line $AE$ touches $\Gamma_3$ at point $m_3 \in m_1
\mu_3$. Denote this line by {\it 3}. Triangles with sides {\it
3'}, {\it 2} and {\it 3}, {\it 1'} are constructed analogously.

\smallskip

There is exactly six tangents from $A$ to conics $\Gamma_1$,
$\Gamma_2$, $\Gamma_3$. We have divided these six lines into two
groups: {\it 1,2,3} i {\it 1',2',3'}. Two tangents enumerated by
different numbers and do not belong to the same group, determine a
Poncelet triangle.

\smallskip

Cubic $\mathcal C$ and the cubic consisting of lines $m_1 M$, $m_2
m_2'$, $m_3 m_3'$ have simple common points $m_1$, $M$, $m_2$,
$m_2'$, $m_3$, $m_3'$, $A$, and point $Z$ as a double one. A
pencil determined by these two cubics contains a curve that passes
through a given point of line $M m_2'$, different from $M, m_2',
m_3'$. This cubic has four common points with line $M m_2'$, so it
decomposes into the line and a conic. Thus, {\it $m_1, m_2, m_3$
are intersection points, different from $A$ and $Z$, of a conic
which contains $A$ and touches cubic $\mathcal C$ at $Z$.}

Converse also holds.

\smallskip

Let an arbitrary conic that contains point $A$ and touches cubic
$\mathcal C$ at $Z$ be given. Denote by $m_1$, $m_2$, $m_3$
remaining intersection points of the curve $\mathcal C$ with this
conic. Each of the lines $m_1Z$, $m_2Z$, $m_3Z$ has another common
point with the cubic $\mathcal C$; denote them by $m_1'$, $m_2'$,
$m_3'$ respectively. By definition of the curve $\mathcal C$, we
have that $m_1,m_1'$; $m_2,m_2'$; $m_3,m_3'$ are pairs of
representative points of some conics $\Gamma_1$, $\Gamma_2$,
$\Gamma_3$ from the pencil $\mathcal F$. Line $Am_1$, besides
being tangent to $\Gamma_1$ at $m_1$, has to touch another conic
from the pencil $\mathcal F$. Take that it is tangent to a conic
$(I)$ at $M$.

\smallskip

Now, in a similar fashion as before, we can conclude that points
$M$, $m_2'$, $m_3'$ are collinear. Applying Lemma 1, it is easily
deduced that conics $\Gamma_1$, $\Gamma_2$, $\Gamma_3$ are joined
to $\Gamma$.

\smallskip

So, we have shown the following: {\it systems of three joined
conics are determined by systems of three intersecting points of
cubic $\mathcal C$ with conics that contain point $A$ and touch
the curve $\mathcal C$ at $Z$}.

\

\noindent{\bf Cayley's Cubic.} Let $D(\lambda)$ be the
discriminant of conic $C + \lambda \Gamma = 0$. We will call the
curve
$$
\mathcal C_0:  Y^2=D(X)
$$
{\sl Cayley's cubic}. Representative points of conic $C + \lambda
\Gamma = 0$ on Cayley's cubic are two points that correspond to
the value $X=\lambda$.

\smallskip

The polar conic of the point $Z$ with respect to cubic $\mathcal
C$ passes through the contact points of the tangents $ZA$, $Z
\alpha$, $Z \beta$, $Z \gamma$ from $Z$ to $\mathcal C$. Thus,
points $\alpha, \beta, \gamma$ are representative points of three
joined conics from the pencil $\mathcal F$. Those three conics are
obviously the decomposable ones. Corresponding values $\lambda$
diminish $D(\lambda)$, and these three representative points on
Cayley's cubic $\mathcal C_0$ lie on the line $Y=0$.

\smallskip

Using the Sylvester's theory of residues, we will show the
following:

\smallskip

{\it Let three representative points of three conics of pencil
$\mathcal F$ be given on the Cayley's cubic $\mathcal C_0$.
Condition for these conics to be joined to the conic $\Gamma$ is
that their representative points are collinear.}

\

\noindent{\bf Sylvester's Theory of Residues.} When considering
algebraic curves of genus 1, like the Cayley's cubic is here,
Abel's theorem can always be replaced by application of this
theory.

\begin{proposition} Let a given cubic and an algebraic
curve of degree $m+n$ meet at $3(m+n)$ points. If there is $3m$
points among them which are placed on a curve of degree $m$, then
the remaining $3n$ points are placed on a curve of degree $n$.
\end{proposition}

If the union of two systems of points is the complete intersection
of a given cubic and some algebraic curve, then we will say that
these two systems are {\it residual} to each other. Now, the
following holds:

\begin{proposition} If systems $\mathcal A$ and $\mathcal
A'$ of points on a given cubic curve have a common residual
system, then they share all residual systems.
\end{proposition}

\begin{proof} Suppose $\mathcal B$ is a system residual to
both $\mathcal A$, $\mathcal A'$ and $\mathcal B'$ is residual to
$\mathcal A$. Then $\mathcal A\cup\mathcal A'$ is residual to
$\mathcal B\cup \mathcal B'$, i.e.\ the system $\mathcal
A\cup\mathcal A'\cup\mathcal A\cup\mathcal A'$ is a complete
intersection of the cubic with an algebraic curve. Since $\mathcal
A\cup\mathcal B$ is also such an intersection, it follows, by the
previous proposition, that $\mathcal A'$ and $\mathcal B'$ are
residual to each other. \end{proof}

Let us note that this proposition can be derived as a consequence
of Abel's theorem, for a plane algebraic curve of arbitrary
degree. However, if the degree is equal to three, i.e.\ the curve
is elliptic, Proposition 2 is equivalent to Abel's theorem.

\

\noindent{\bf Condition for Existence of a Poncelet Polygon.} Let
conics $\Gamma, \Gamma_1, \dots, \Gamma_n$ be from a pencil. If
there exists a polygon inscribed in $\Gamma$ and circumscribed
about $\Gamma_1, \dots, \Gamma_n$, we are going to say that conics
$\Gamma_1, \dots, \Gamma_n$ are {\sl joined} to $\Gamma$. Then,
similarly as in the case of the triangle, it can be proved that
tangents from the point $A \in \Gamma$ to $\Gamma_1, \dots,
\Gamma_n$ can be divided into two groups such that any Poncelet
$n$-polygon with vertex $A$ has exactly one side in each of the
groups.

\smallskip

This division of tangents gives a division of characteristic
points of conics $\Gamma_1$, \dots, $\Gamma_n$ into two groups on
$\mathcal C$ and, therefore, a division into two groups on
Cayley's cubic $\mathcal C_0$: {\it 1,2,3,} \dots and {\it
1',2',3',} \dots.

\smallskip

Let $ABCD \dots KL$ be a Poncelet polygon, and let $(AC), (AD),
\dots$ be conics determined like in the definition of a
circumscribed polygon. Let $c, \gamma$; $d, \delta$; \dots be a
corresponding characteristic points on $\mathcal C_0$, such that
triples {\it 1, 2}, $c$; $\gamma$, {\it 3}, $d$; $\delta$, {\it
4}, $e$ are characteristic points of the same group with respect
to corresponding conics.

\smallskip

Points {\it 1, 2}, $c$ are collinear, as $\gamma$, {\it 3}, $d$
are. Thus, {\it 1, 2, 3}, $d$ are residual with $c, \gamma$. Line
$c \gamma$ contains point $Z$, so system $c, \gamma$ is residual
with $Z$, too. It is possible to show that $Z$ is a triple point
of curve $\mathcal C_0$ and it follows that it is residual with
system $Z,Z$. This implies that points {\it 1,2,3}, $d,Z,Z$ are
placed on a conic.

\smallskip

If we take a coordinate system such that the tangent line to
$\mathcal C_0$ at $Z$ is the infinite line and the axis $Oy$ is
line $AZ$, we will have:

four conics are joined to $\Gamma$ if and only if their
characteristic points of the same group are on a parabola with the
asymptotic direction $Oy$.

\smallskip

Continuing deduction in the same manner, we can conclude: {\it $3n
- p$ points of the cubic $\mathcal C_0$ are characteristic points
of same group for $3n - p$ conics $(1 \leq p \leq 3)$ joined to
$\Gamma$, if and only if these points are placed on a curve of
degree $n$ which has $Oy$ as an asymptotic line of the order $p$.}

\

\noindent{\bf Cayley's Condition.} Let $\mathcal C_0 : y^2 = D(x)$
be the Cayley's cubic, where $D(x)$ is the discriminant of the
conic $C + x \Gamma = 0$ from pencil $\mathcal F$. A system of $n$
conics joined to $\Gamma$ is determined by $n$ values $x$ if and
only if these $n$ values are abscissae of intersecting points of
$\mathcal C_0$ and some algebraic curve. Plugging $D(x)$ instead
of $y^2$ in the equation of this curve, we obtain:
$$
P(x)y + Q(x) = 0,
$$
that is
$$
P(x)\sqrt{D(x)} + Q(x) = 0.
$$
From there:
$$
\sqrt{D(x)}(a_0 x^{p-2} + a_1 x^{p-3} + \dots + a_{p-2}) +
    (b_0 x^p + b_1 x^{p-1} + \dots + b_p) = 0, \quad n = 2p;
$$
$$
\sqrt{D(x)}(a_0 x^{p-1} + a_1 x^{p-2} + \dots + a_{p-1}) +
    (b_0 x^p + b_1 x^{p-1} + \dots + b_p) = 0, \quad n = 2p+1.
$$

If $\lambda_1,\dots,\lambda_k$ denote parameters corresponding to
$\Gamma_1,\dots,\Gamma_k$ respectively, then existence of a
Poncelet polygon inscribed in $\Gamma$ and circumscribed about
$\Gamma_1,\dots,\Gamma_k$ is equivalent to:
$$
\left|\begin{array}{ccccccccc} 1 & \lambda_1 & \lambda_1^2 & \dots
& \lambda_1^p & \sqrt{D(\lambda_1)} & \lambda_1\sqrt{D(\lambda_1)}
& \dots &
\lambda_1^{p-2}\sqrt{D(\lambda_1)}\\
\dots\\
\dots\\
1 & \lambda_k & \lambda_k^2 & \dots & \lambda_k^p &
\sqrt{D(\lambda_k)} & \lambda_k\sqrt{D(\lambda_k)} & \dots &
\lambda_k^{p-2}\sqrt{D(\lambda_k)}
\end{array}\right|=0,
$$
for $k=2p$;
$$
\left|\begin{array}{ccccccccc} 1 & \lambda_1 & \lambda_1^2 & \dots
& \lambda_1^p & \sqrt{D(\lambda_1)} & \lambda_1\sqrt{D(\lambda_1)}
& \dots &
\lambda_1^{p-1}\sqrt{D(\lambda_1)}\\
\dots\\
\dots\\
1 & \lambda_k & \lambda_k^2 & \dots & \lambda_k^p &
\sqrt{D(\lambda_k)} & \lambda_k\sqrt{D(\lambda_k)} & \dots &
\lambda_k^{p-1}\sqrt{D(\lambda_k)}
\end{array}\right|=0,
$$
for $k=2p+1$.

\smallskip

There exists an $n$-polygon inscribed in $\Gamma$ and
circumscribed about $C$ if and only if it is possible to find
coefficients $a_0$, $a_1$, \dots; $b_0$, $b_1$, \dots such that
function $P(x) \sqrt{D(x)} + Q(x)$ has $x=0$ as a root of the
multiplicity $n$.

\smallskip

For $n=2p$, this is equivalent to the existence of a non-trivial
solution of the following system:
$$
\begin{array}{ccccccccc}
    a_0 C_3     & + & a_1 C_4     & + & \dots & + & a_{p-2} C_{ p+1} & = & 0 \\
    a_0 C_4     & + & a_1 C_5     & + & \dots & + & a_{p-2} C_{ p+2} & = & 0 \\
    \dots  \\
    a_0 C_{p+1} & + & a_1 C_{p+2} & + & \dots & + & a_{p-2} C_{2p-1} & = & 0,
\end{array}
$$
where
$$ \sqrt{D(x)} = A + Bx + C_2 x^2 + C_3 x^3 + \dots.$$
Finally, for $n=2p$, we obtain the Cayley's condition
$$ \left | \begin{array}{llll}
     C_3     & C_4     & \dots & C_{p+1} \\
     C_4     & C_5     & \dots & C_{p+2} \\
      & & \dots                          \\
     C_{p+1} & C_{p+2} & \dots & C_{2p-1}
     \end{array} \right |=0.$$
Similarly, for $n = 2p+1$, we obtain:
$$ \left | \begin{array}{llll}
     C_2     & C_3     & \dots & C_{p+1} \\
     C_3     & C_4     & \dots & C_{p+2} \\
      & & \dots                         \\
     C_{p+1} & C_{p+2} & \dots & C_{2p}
     \end{array} \right |=0.
$$

These results can be directly applied to the billiard system
within an ellipse: to determine whether a billiard trajectory with
a given confocal caustic is periodic, we need to consider the
pencil determined by the boundary and the caustic curve.

\subsection{Some Applications of Lebesgue's Results}

Now, we are going to apply the Lebesgue's results to billiards
systems within several confocal conics in the plane.

\smallskip

Consider the dual plane. The case with two ellipses, when the
billiard trajectory is placed between them and particle bounces to
one and another of them alternately, is of special interest.

\begin{corollary} The condition for the existence of
$2m$-periodic billiard trajectory which bounces exactly $m$ times
to the ellipse $\Gamma_1^*=C^*$ and $m$ times to
$\Gamma_2^*=(C+\gamma\Gamma)^*$, having $\Gamma^*$ for the
caustic, is:
$$
\det\left(\begin{array}{llll}
f_0(0) & f_1(0) &\dots & f_{2m-1}(0)\\
f_0'(0) & f_1'(0) &\dots & f_{2m-1}'(0)\\
 & & \dots\\
f_0^{(m-1)}(0) & f_1^{(m-1)}(0) &\dots & f_{2m-1}^{(m-1)}(0)\\
f_0(\gamma) & f_1(\gamma) &\dots & f_{2m-1}(\gamma)\\
f_0'(\gamma) & f_1'(\gamma) &\dots & f_{2m-1}'(\gamma)\\
 & & \dots\\
f_0^{(m-1)}(\gamma) & f_1^{(m-1)}(\gamma)&\dots &
f_{2m-1}^{(m-1)}(\gamma)
\end{array}\right)=0,
$$
where $f_j=x^j$, $(0\le j\le m)$, $f_{m+i}=x^{i-1}\sqrt{D(x)}$,
$(1\le i\le m-1)$.
\end{corollary}

We consider a simple example with four bounces on each of the two
conics.

\begin{example}\label{ex:4alt}
{\rm The condition on a billiard trajectory placed between
ellipses $\Gamma_1^*$ and $\Gamma_2^*$, to be closed after 4
alternate bounces to each of them is:
$$
\det X = 0,
$$
where the elements of the $3\times 3$ matrix $X$ are:
$$
\aligned
X_{11}&= -4B_0+B_1\gamma+4C_0+3C_1\gamma+2C_2\gamma^2+C_3\gamma^3 \\
X_{12}&=-3B_0+B_1\gamma+3C_0+2C_1\gamma+C_2\gamma^2\\
X_{13}&=-2B_0+B_1\gamma+2C_0+C_1\gamma \\
X_{21}&=-6B_0+B_2\gamma^2+6C_0+6C_1\gamma+4C_2\gamma^2+3C_3\gamma^3\\
X_{22}&=-6B_0+B_1\gamma+B_2\gamma^2+6C_0+4C_1\gamma+3C_2\gamma^2 \\
X_{23}&=-5B_0+2B_1\gamma+B_2\gamma^2+5C_0+3C_1\gamma \\
X_{31}&=-4B_0+B_3\gamma^3+4C_0+4C_1\gamma+4C_2\gamma^2+3C_3\gamma^3\\
X_{32}&=-4B_0+B_2\gamma^2+B_3\gamma^3+4C_0+4C_1\gamma+3C_2\gamma^2\\
X_{33}&=-4B_0+B_1\gamma+B_2\gamma^2+B_3\gamma^3+4C_0+3C_1\gamma,
\endaligned
$$
with $C_i, B_i$ being coefficients in the Taylor expansions around
$x=0$ and $x=\gamma$ respectively:
$$
\aligned
\sqrt{D(x)}&=C_0+C_1x+C_2x^2+\dots,\\
\sqrt{D(x)}&=B_0+B_1(x-\gamma)+B_2(x-\gamma)^2+\dots.
\endaligned
$$

On Figure \ref{fig:4alt}, we see a Poncelet octagon inscribed in
$\Gamma$ and circumscribed about $\Gamma_1$ and $\Gamma_2$. In the
dual plane, the billiard trajectory that corresponds to this
octagon, has the dual conic $\Gamma^*$ as the caustic (see Figure
\ref{fig:4alt.dual}).}
\begin{figure}[h]
\centering
\begin{minipage}[t]{0.44\textwidth}
\centering
\includegraphics[width=5cm,height=4cm]{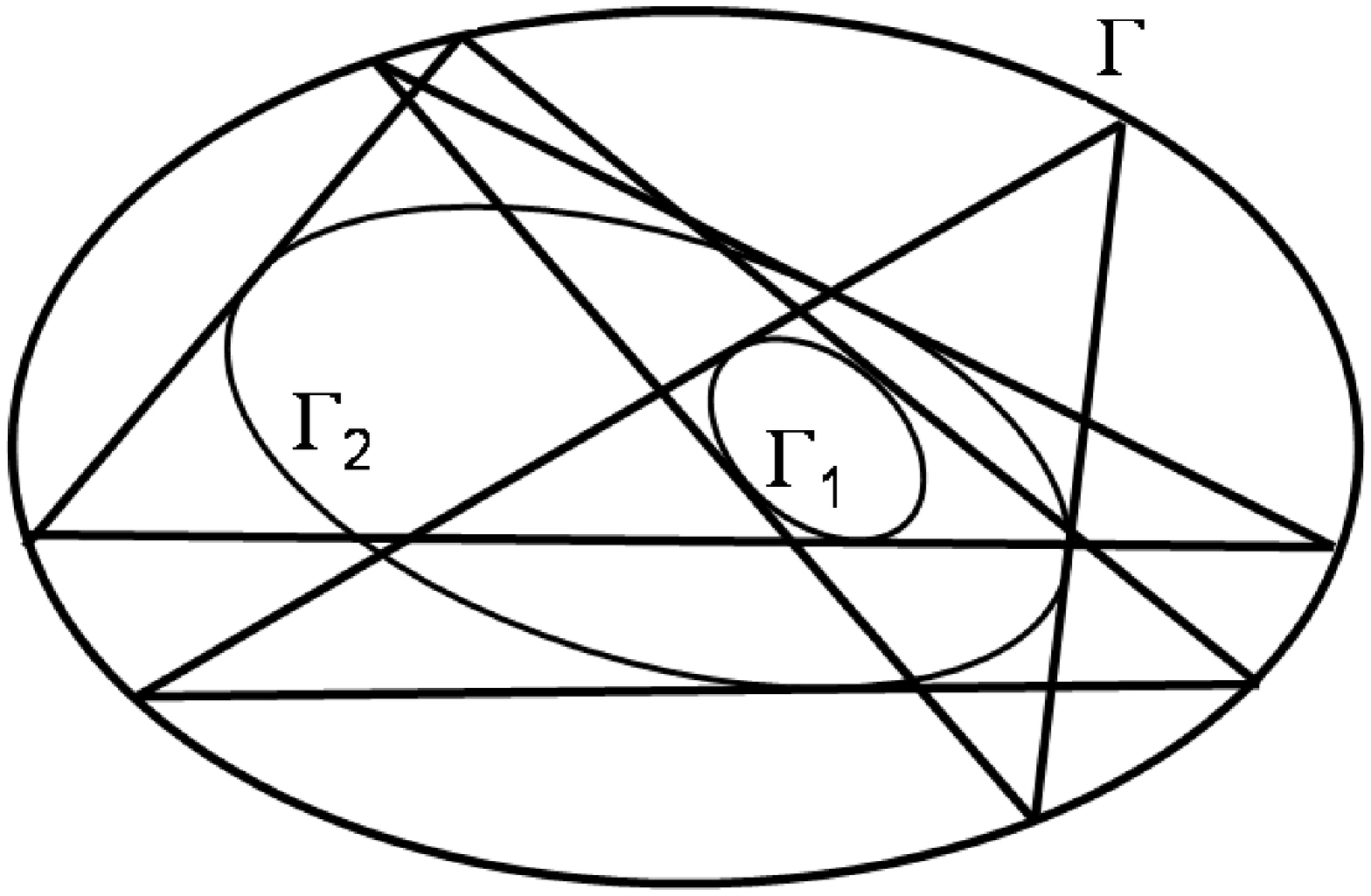}
\parbox{5cm}{\caption{A Poncelet octagon whose sides touch the conics
$\Gamma_1$ and $\Gamma_2$ alternately}\label{fig:4alt}}
\end{minipage}
\begin{minipage}[t]{0.44\textwidth}
\centering
\includegraphics[width=5cm,height=4cm]{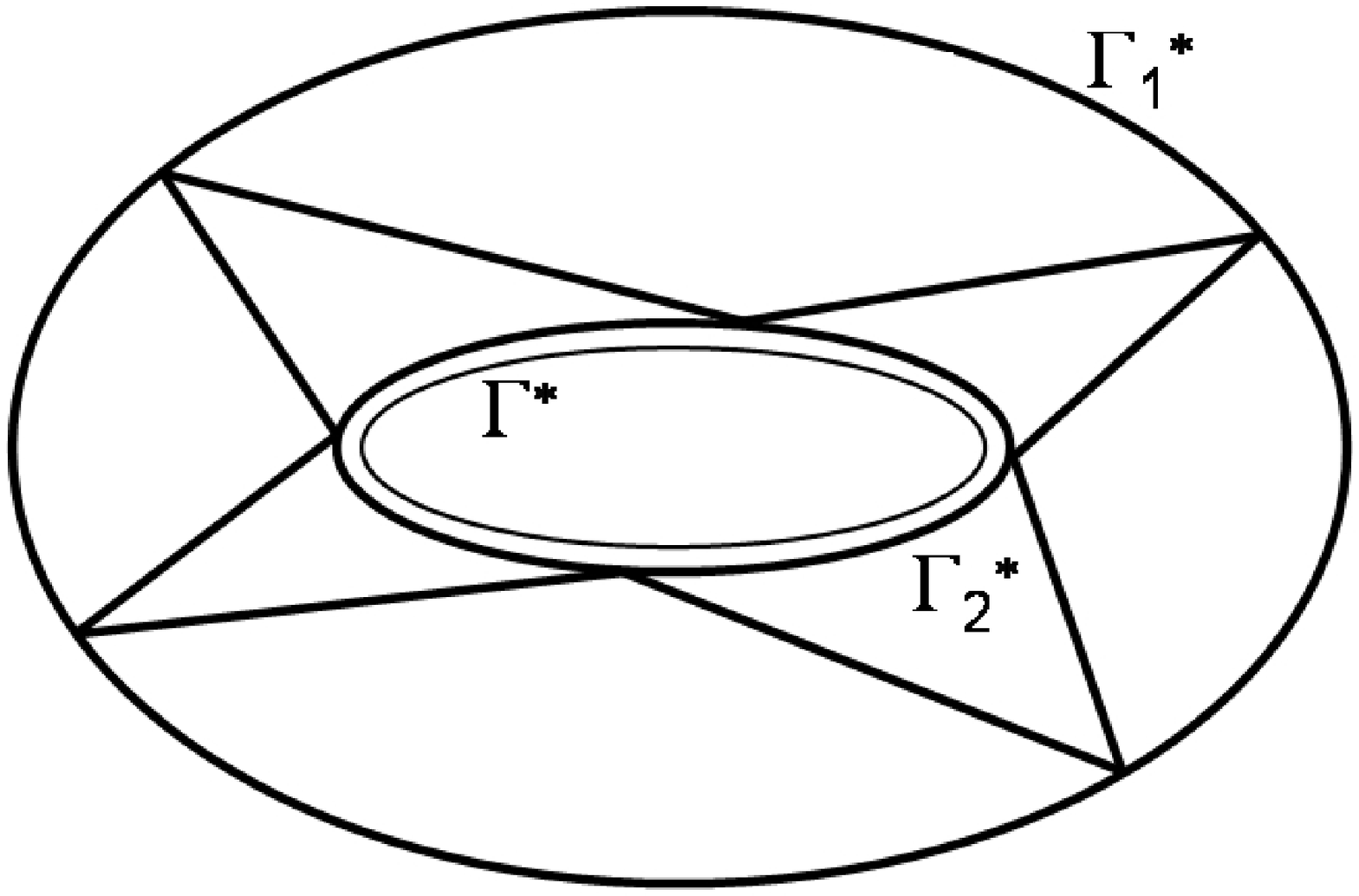}
\parbox{5cm}{\caption{A closed billiard trajectory in the domain bounded by
two confocal ellipses}\label{fig:4alt.dual}}
\end{minipage}
\end{figure}
\end{example}

\section{Points of Finite Order on the Jacobian of a Hyperelliptic
Curve}\label{hyper}

In order to prepare the algebro-geometric background for the rest
of the article, in this section we are going to give the
analytical characterization of some classes of finite order
divisors on a hyperelliptic curve.

\smallskip

Let the curve $\mathcal C$ be given by
$$
y^2=(x-x_1)\dots(x-x_{2g+1}), \quad x_i\neq x_j \ \ \text{when}\ \
i\neq j.
$$
It is a regular hyperelliptic curve of genus $g$, embedded in
$P^2$. Let $\mathcal J(\mathcal C)$ be its Jacobian variety and
$$
\mathcal A\ :\ \mathcal C \rightarrow \mathcal J(\mathcal C)
$$
the Abel-Jacobi map.

\smallskip

Take $E$ to be the point which corresponds to the value
$x=\infty$, and choose $\mathcal A(E)$ to be the neutral in
$\mathcal J(\mathcal C)$. According to the Abel's theorem
\cite{Gun}, $\mathcal A(P_1)+\dots+\mathcal A(P_n)=0$ if and only
if there exists a meromorphic function $f$ with zeroes
$P_1,\dots,P_n$ and a pole of order $n$ at the point $E$. Let
$\mathcal L(nE)$ be the vector space of meromorphic functions on
$\mathcal C$ with a unique pole $E$ of order at most $n$, and
$f_1,\dots,f_k$ a basis of $\mathcal L(nE)$. The mapping
$$
F:\mathcal C\to P^{k-1}, \quad X\mapsto[f_1(X),\dots,f_k(X)]
$$
is a projective embedding whose image is a smooth algebraic curve
of degree $n$. Hyperplane sections of this curve are zeroes of
functions from $\mathcal L(nE)$. Thus, the equality $n\mathcal
A(P)=0$ is equivalent to:
\begin{equation}\label{rank.f}
\rank\left(\begin{array}{llll}
f_1(P) & f_2(P) & \dots & f_k(P) \\
f_1'(P) & f_2'(P) & \dots & f_k'(P) \\
 & & \dots \\
f_1^{(n-1)}(P) & f_2^{(n-1)}(P) & \dots & f_k^{(n-1)}(P)
\end{array}\right)<k.
\end{equation}

\begin{lemma} For $n\le2g$, there does not exist a point $P$ on
the curve $\mathcal C$, such that $n\mathcal A(P)=0$ and $P\neq
E$.
\end{lemma}

\begin{proof} Let $P$ be a point on $\mathcal C$, $P\neq
E$, and $x=x_0$ its corresponding value. Consider the case of $n$
even. Since $E$ is a branch point of a hyperelliptic curve, its
Weierstrass gap sequence is $1,3,5,\dots,2g-1$ \cite{Gun}. Now,
applying the Riemann-Roch theorem, we obtain $\dim \mathcal
L(nE)=n/2+1$. Choosing a basis $1,x,\dots,x^{n/2}$ for $\mathcal
L(nE)$, and substituting in (\ref{rank.f}), we come to a
contradiction. \end{proof}

\begin{lemma}\label{nP=0} Let $P(x_0,y_0)$ be a non-branching point
on the curve $\mathcal C$. For $n>2g$, equality $n\mathcal A(P)=0$
is equivalent to:
\begin{equation}\label{rank.B}
\rank\left(\begin{array}{llll}
B_{m+1} & B_m & \dots & B_{g+2}\\
B_{m+2} & B_{m+1} & \dots & B_{g+3}\\
 & & \dots\\
B_{2m-1} & B_{2m-2} & \dots & B_{m+g}
\end{array}\right)<m-g,
\quad \text{when} \ \ n=2m,
\end{equation}
$$
\rank\left(\begin{array}{llll}
B_{m+1} & B_m & \dots & B_{g+1}\\
B_{m+2} & B_{m+1} & \dots & B_{g+2}\\
 & & \dots\\
B_{2m} & B_{2m-1} & \dots & B_{m+g}
\end{array}\right)<m-g+1,
\quad \text{when} \ \ n=2m+1,
$$
and
$\sqrt{(x-x_1)\dots(x-x_{2g+1})}=B_0+B_1(x-x_0)+B_2(x-x_0)^2+B_3(x-x_0)^3+\cdots$.
\end{lemma}

\begin{proof} This claim follows from previous results by
choosing a basis for $\mathcal L(nE)$:
$$
\displaylines{ 1,x,\dots,x^m,y,xy,\dots,x^{m-g-1}y\ \text{if}\
n=2m,\cr 1,x,\dots,x^m,y,xy,\dots,x^{m-g}y\ \text{if}\ n=2m+1, }
$$
 similarly as in
\cite{GH2}. \end{proof}

In the next lemma, we are going to consider the case when the
curve $\mathcal C$ is singular, i.e.\ when some of the values
$x_1$, $x_2$, \dots, $x_{2g+1}$ coincide.

\begin{lemma}\label{sing.kriva} Let the curve $\mathcal C$ be given by
$$
y^2=(x-x_1)\dots(x-x_{2g+1}),\quad x_1\cdot x_2\cdot\dots\cdot
x_{2g+1}\neq0,
$$
$P_0$ one of the points corresponding to the value $x=0$ and $E$
the infinite point on $\mathcal C$. Then $2n P_0\sim 2nE$ is
equivalent to {\rm (\ref{rank.B})}, where
$$
y=\sqrt{(x-x_1)\dots(x-x_{2g+1})}=B_0+B_1x+B_2x^2+\dots
$$
is the Taylor expansion around the point $P_0$.
\end{lemma}

\begin{proof} Suppose that, among $x_1, \dots, x_{2g+1}$,
only $x_{2g}$ and $x_{2g+1}$ have same values. Then $(x_{2g},0)$
is an ordinary double point on $\mathcal C$. The normalization of
the curve $\mathcal C$ is the pair $(\tilde{\mathcal{C}},\pi)$,
where $\tilde{\mathcal{C}}$ is the curve given by:
$$
\tilde{\mathcal{C}}: \tilde y^2=(\tilde x-x_1)\dots(\tilde
x-x_{2g-1}),
$$
and $\pi:\tilde{\mathcal{C}}\to\mathcal C$ is the projection:
$$
(\tilde x,\tilde y) \buildrel\pi\over\longmapsto (x=\tilde x,\
y=(\tilde x-x_{2g})\tilde y).
$$
The genus of $\tilde{\mathcal C}$ is $g-1$.

\smallskip

Denote by $A$ and $B$ points on $\tilde{\mathcal C}$ which are
mapped to the singular point $(x_{2g},0)\in\mathcal C$ by the
projection $\pi$. Any other point on $\mathcal C$ is the image of
a unique point of the curve $\tilde{\mathcal C}$. Let
$$
\pi(\tilde E)=E,\quad \pi(\tilde P_0)=P_0.
$$

The relation $2nP_0\sim 2nE$ holds if and only if there exists a
meromorphic function $f$ on $\tilde{\mathcal C}$, $f\in \mathcal
L(2n\tilde E)$, having a zero of order $2n$ at $\tilde P_0$ and
satisfying $f(A)=f(B)$.

\smallskip

For $n\le g-1$, according to Lemma \ref{nP=0}, $2n\tilde E \sim
2n\tilde P_0$ cannot hold. For $n\ge g$, choose the following
basis of the space $\mathcal L(2n\tilde E)$:
$$
1,\ \tilde y,\ f_1 \circ \pi,\ \dots,\ f_{n-g-1}\circ\pi,
$$
where $1, f_1, \dots, f_{n-g-1}$ is a basis of $\mathcal L(2nE)$
as in the proof of Lemma \ref{nP=0}.

\smallskip

Since $\tilde y$ is the only function in the basis which has
different values at points $A$ and $B$, we obtain that the
condition
$$
2n\tilde E\sim 2n\tilde P_0
$$
is equivalent to (\ref{rank.B}).

\smallskip

Cases when $\mathcal C$ has more singular points or singularities
of higher order, can be discussed in the same way.
\end{proof}

\begin{lemma} Let the curve $\mathcal C$ be given by
$$
y^2=(x-x_1)\dots(x-x_{2g+2}),
$$
with all $x_i$ distinct from $0$, and $Q_+$, $Q_-$ the two points
on $\mathcal C$ over the point $x=0$. Then $nQ_+\sim nQ_-$ is
equivalent to:
\begin{equation}\label{rank.B2}
\rank\left(\begin{array}{llll}
B_{g+2} & B_{g+3} & \dots & B_{n+1} \\
B_{g+3} & B_{g+4} & \dots & B_{n+2} \\
\dots & \dots & \dots & \dots \\
\dots & \dots & \dots & \dots \\
\\
B_{g+n} & \dots & \dots & B_{2n-1}
\end{array}\right)<n-g
\quad \text{and} \quad n>g,
\end{equation}
where $y=\sqrt{(x-x_1)\dots(x-x_{2g+2})}=B_0+B_1x+B_2x^2+\dots$ is
the Taylor expansion around the point $Q_-$.
\end{lemma}

\begin{proof} $\mathcal C$ is a hyperelliptic curve of
genus $g$. The relation $nQ_+\equiv nQ_-$ means that there exists
a meromorphic function on $\mathcal C$ with a pole of order $n$ at
the point $Q_+$, a zero of the same order at $Q_-$ and neither
other zeros nor poles. Denote by $L(nQ_+)$ the vector space of
meromorphic functions on $\mathcal C$ with a unique pole $Q_+$ of
order at most $n$. Since $Q_+$ is not a branching point on the
curve, $\dim L(nQ_+)=1$ for $n\le g$, and $\dim L(nQ_+)=n-g+1$,
for $n>g$. In the case $n\le g$, the space $L(nQ_+)$ contains only
constant functions, and the divisors $nQ_+$ and $nQ_-$ can not be
equivalent. If $n\ge g+1$, we choose the following basis for
$L(nQ_+)$:
$$
1, f_1, \dots, f_{n-g},
$$
where
$$
f_k=\frac{y-B_0-B_1x-\dots-B_{g+k-1}x^{g+k-1}}{x^{g+k}}.
$$

Thus, $nQ_+\equiv nQ_-$ if there is a function $f\in L(nQ_+)$ with
a zero of order $n$ at $Q_-$, i.e., if there exist constants
$\alpha_0, \dots, \alpha_{n-g}$, not all equal to 0, such that:
$$
\begin{array}{ccccccccl}
\alpha_0 & + & \alpha_1 f_1(Q_-) & + & \dots & + & \alpha_{n-g}f_{n-g}(Q_-) & = & 0 \\
 &  & \alpha_1 f_1'(Q_-) & + & \dots & + & \alpha_{n-g}f_{n-g}'(Q_-) & = & 0 \\
 &   \dots \\
 &   \dots \\
\\
 &  & \alpha_1 f_1^{(n-1)}(Q_-) & + & \dots & + & \alpha_{n-g}f_{n-g}^{(n-1)}(Q_-) & = & 0.
\end{array}
$$
Existence of a non-trivial solution to this system of linear
equations is equivalent to the condition (\ref{rank.B2}).

\smallskip

When some of the values $x_1,\dots,x_{2g+2}$ coincide, the curve
$\mathcal C$ is singular. This case can be considered by the
procedure of normalization of the curve, as in Lemma 4. The
condition for the equivalence of the divisors $nQ_+$ i $nQ_-$, in
the case when $\mathcal C$ is singular, is again (\ref{rank.B2}).
\end{proof}

\section{Periodic Billiard Trajectories inside $k$ Confocal
Quadrics in $\mathbf R^{\mathbf d}$}\label{k.quad}

\

Darboux was the first who considered a higher-dimensional
generalization of Poncelet theorem. Namely, he investigated
light-rays in the three-dimensional case ($d=3$) and announced the
corresponding complete Poncelet theorem in \cite{Dar1} in 1870.

\smallskip

Higher-dimensional generalizations of CPT ($d\ge3$) were obtained
quite recently in \cite{CCS}, and the related Cayley-type
conditions were derived by the authors \cite{DR3}.

\smallskip

The main goal of this section is to present detailed proof of
Cayley-type condition for generalized CPT, together with
discussions and examples.

\smallskip

Consider an ellipsoid in ${\mathbb R}^d$:
$$
\frac {x_1^2}{a_1}+\dots + \frac {x_d^2}{a_d}=1,\quad
a_1>\dotsb>a_d>0,
$$
and the related system of Jacobian elliptic coordinates
$(\lambda_1,\dots, \lambda_d)$ ordered by the condition
$$
\lambda_1>\lambda_2>\dotsb> \lambda_d.
$$
If we denote:
$$
Q_{\lambda}(x)=\frac {x_1^2}{a_1-\lambda}+\dots + \frac
{x_d^2}{a_d-\lambda},
$$
then any quadric from the corresponding confocal family is given
by the equation of the form:
\begin{equation}\label{konfokalna.familija}
\mathcal Q_{\lambda}:\ Q_{\lambda}(x) = 1.
\end{equation}

The famous Chasles theorem states that any line in the space
$\mathbb R^d$ is tangent to exactly $d-1$ quadrics from a given
confocal family. Next lemma gives an important condition on these
quadrics.

\begin{lemma} Suppose a line $\ell$ is tangent to quadrics
$\mathcal Q_{\alpha_1},\dots,\mathcal Q_{\alpha_{d-1}}$ from the
family {\rm (\ref{konfokalna.familija})}. Then Jacobian
coordinates $(\lambda_1,\dots, \lambda_d)$ of any point on $\ell$
satisfy the inequalities $\mathcal P(\lambda_s)\ge 0$,
$s=1,\dots,d$, where
$$
\mathcal
P(x)=(a_1-x)\dots(a_d-x)(\alpha_1-x)\dots(\alpha_{d-1}-x).
$$
\end{lemma}

\begin{proof} Let $x$ be a point of $\ell$,
$(\lambda_1,\dots,\lambda_d)$ its Jacobian coordinates, and $y$ a
vector parallel to $\ell$. The equation $Q_{\lambda}(x+ty)=1$ is
quadratic with respect to $t$. Its discriminant is:
$$
\Phi_{\lambda}(x,y) =
Q_{\lambda}(x,y)^2-Q_{\lambda}(y)\bigl(Q_{\lambda}(x)-1\bigr),
$$
where
$$
Q_{\lambda}(x,y) = \frac{x_1y_1}{a_1-\lambda}+\dots +
\frac{x_dy_d}{a_d-\lambda}.
$$
By \cite{Mo},
$$
\Phi_{\lambda}(x,y)=\frac
{(\alpha_1-\lambda)\dots(\alpha_{d-1}-\lambda)}
{(a_1-\lambda)\dots(a_d-\lambda)}.
$$
For each of the coordinates $\lambda=\lambda_s$, ($1\le s\le d$),
the quadratic equation has a solution $t=0$; thus, the
corresponding discriminants are non-negative. This is obviously
equivalent to $\mathcal P(\lambda_s)\ge0$.
\end{proof}

\subsection{Billiard inside a Domain Bounded by Confocal
Qua\-drics}

Suppose that a bounded domain $\Omega\subset\mathbb R^d$ is given
such that its boundary $\partial\Omega$ lies in the union of
several quadrics from the family (\ref{konfokalna.familija}).
Then, in elliptic coordinates, $\Omega$ is given by:
$$
\beta_1'\le\lambda_1\le\beta_1'', \quad\dots,\quad
\beta_d'\le\lambda_d\le\beta_d'',
$$
where $a_{s+1}\le\beta_s'<\beta_s''\le a_s$ for $1\le s\le d-1$
and $-\infty<\beta_d'<\beta_d''\le a_d$.

\smallskip

Consider a billiard system within $\Omega$ and let $\mathcal
Q_{\alpha_1}$, \dots, $\mathcal Q_{\alpha_{d-1}}$ be caustics of
one of its trajectories. For any $s=1,\dots, d$, the set
$\Lambda_s$ of all values taken by the coordinate $\lambda_s$ on
the trajectory is, according to Lemma 6, included in
$\Lambda_s'=\{\, \lambda\in[\beta_s',\beta_s'']\, :\, \mathcal
P(\lambda)\ge0\,\}$. By \cite{Kn}, each of the intervals
$(a_{s+1}, a_s)$, $(2\le s\le d)$ contains at most two of the
values $\alpha_1,\dots,\alpha_{d-1}$, the interval $(-\infty,a_d)$
contains at most one of them, while none is included in
$(a_1,+\infty)$. Thus, for each $s$, the following three cases are
possible:

\smallskip

\noindent{\it First case:}
$\alpha_i,\alpha_j\in[\beta_s',\beta_s'']$, $\alpha_i<\alpha_j$.
Since any line which contains a segment of the trajectory touches
$\mathcal Q_{\alpha_i}$ and $\mathcal Q_{\alpha_j}$, the whole
trajectory is placed between these two quadrics. The elliptic
coordinate $\lambda_s$ has critical values at points where the
trajectory touches one them, and remains monotonous elsewhere.
Hence, meeting points with with $\mathcal Q_{\alpha_i}$ and
$\mathcal Q_{\alpha_j}$ are placed alternately along the
trajectory and $\Lambda_s=\Lambda_s'=[\alpha_i,\alpha_j]$.

\smallskip

\noindent{\it Second case:} Among $\alpha_1,\dots,\alpha_{d-1}$,
only $\alpha_i$ is in $[\beta_s', \beta_s'']$. $\mathcal P$ is
non-negative in exactly one of the intervals: $[a_{s+1},
\alpha_i]$, $[\alpha_i,a_s]$, let us take in the first one. Then
the trajectory has bounces only on $\mathcal Q_{\beta_s'}$. If
$\alpha_i\neq\beta_s''$, the billiard particle never reaches the
boundary $\mathcal Q_{\beta_s''}$. The coordinate $\lambda_s$ has
critical values at meeting points with $\mathcal Q_{\beta_s'}$ and
the caustic $\mathcal Q_{\alpha_i}$, and remains monotonous
elsewhere. Hence, $\Lambda_s=\Lambda_s'=[\beta_s',\alpha_i]$. If
$\mathcal P$ is non-negative in $[\alpha_i,a_s]$, then we obtain
$\Lambda_s=\Lambda_s'=[\alpha_i,\beta_s'']$.

\smallskip

\noindent{\it Third case:} The segment $[\beta_s',\beta_s'']$ does
not contain any of values $\alpha_1$, \dots, $\alpha_{d-1}$. Then
$\mathcal P$ is non-negative in $[\beta_s',\beta_s'']$. The
coordinate $\lambda_s$ has critical values only at meeting points
with boundary quadrics $\mathcal Q_{\beta_s'}$ and $\mathcal
Q_{\beta_s''}$, and changes mo\-no\-to\-nously between them. This
implies that the billiard particle bounces of them alternately.
Obviously, $\Lambda_s=\Lambda_s'=[\beta_s',\beta_s'']$.

\smallskip

Denote $[\gamma_s',\gamma_s'']:=\Lambda_s=\Lambda_s'$. Notice that
the trajectory meets quadrics of any pair $\mathcal
Q_{\gamma_s'}$, $\mathcal Q_{\gamma_s''}$ alternately. Thus, any
periodic trajectory has the same number of intersection points
with each of them.

\smallskip

Let us make a few remarks on the case when $\gamma_s'=a_{s+1}$ or
$\gamma_s''=a_s$. This means that either a part of
$\partial\Omega$ is a degenerate quadric from the confocal family
or $\Omega$ is not bounded, from one side at least, by a quadric
of the corresponding type. Discussion of the case when $\Omega$ is
bounded by a coordinate hyperplane does not differ from the one we
have just made. On the other hand, non-existence of a part of the
boundary means that the coordinate $\lambda_s$ will have extreme
values at the points of intersection of the trajectory with the
corresponding hyperplane. Since a closed trajectory intersects any
hyperplane even number of times, it follows that the coordinate
$\lambda_s$ is taking each of its extreme values even number of
times during the period.

\begin{theorem}\label{uslov.k} A trajectory of the billiard system
within $\Omega$ with caustics $\mathcal Q_{\alpha_1}$, \dots,
$\mathcal Q_{\alpha_{d-1}}$ is periodic with exactly $n_s$ points
at $\mathcal Q_{\gamma_s'}$ and $n_s$ points at $\mathcal
Q_{\gamma_s''}$ $(1\le s\le d)$ if and only if
\begin{equation*}\label{uslov}
\sum_{s=1}^d n_s\left(\mathcal A(P_{\gamma_s'}) -\mathcal
A(P_{\gamma_s''})\right)=0
\end{equation*}
on the Jacobian of the curve
$$
\Gamma \ :\ y^2=\mathcal P(x):=
(a_1-x)\cdots(a_d-x)(\alpha_1-x)\cdots(\alpha_{d-1}-x).
$$
Here, $\mathcal A$ denotes the Abel-Jacobi map, where
$P_{\gamma_s'}$, $P_{\gamma_s''}$ are points on $\Gamma$ with
coordinates $P_{\gamma_s'}=\left(\gamma_s', (-1)^s \sqrt {\mathcal
P(\gamma_s')}\right)$, $P_{\gamma_s''}=\left(\gamma_s'', (-1)^s
\sqrt {\mathcal P(\gamma_s'')}\right)$.
\end{theorem}

\begin{proof} Following Jacobi \cite{Jac} and Darboux
\cite{Dar3}, let us consider the equations:
\begin{equation}\label{sistem}
\sum_{s=1}^d\frac{d\lambda_s}{\sqrt{\mathcal P(\lambda_s)}}=0,
\quad \sum_{s=1}^d\frac{\lambda_s d\lambda_s}{\sqrt{\mathcal
P(\lambda_s)}}=0, \quad \dots, \quad
\sum_{s=1}^d\frac{\lambda_s^{d-2}d\lambda_s}{\sqrt{\mathcal
P(\lambda_s)}}=0,
\end{equation}
where, for any fixed $s$, the square root $\sqrt{\mathcal
P(\lambda_s)}$ is taken with the same sign in all of the
expressions. Then (\ref{sistem}) represents a system of differential
equations of a line tangent to $\mathcal Q_{\alpha_1}$, \dots,
$\mathcal Q_{\alpha_{d-1}}$. Besides that,
\begin{equation}\label{duzina}
\sum_{s=1}^d\frac{\lambda_s^{d-1} d\lambda_s}{\sqrt{\mathcal
P(\lambda_s)}}=2d\ell,
\end{equation}
 where $d\ell$ is the element
of the line length.

\smallskip

Attributing all possible combinations of signs to $\sqrt{\mathcal
P(\lambda_1)}$, \dots, $\sqrt{\mathcal P(\lambda_d)}$, we can
obtain $2^{d-1}$ non-equivalent systems (\ref{sistem}), which
correspond to $2^{d-1}$ different tangent lines to $\mathcal
Q_{\alpha_1}$, \dots, $\mathcal Q_{\alpha_{d-1}}$ from a generic
point of the space. Moreover, the systems corresponding to a line
and its reflection to a given hyper-surface $\lambda_s=\const$
differ from each other only in signs of the roots $\sqrt{\mathcal
P(\lambda_s)}$.

\smallskip

Solving (\ref{sistem}) and (\ref{duzina}) as a system of linear
equations with respect to $\dfrac{d\lambda_s}{\sqrt{\mathcal
P(\lambda_s)}}$, we obtain:
$$
\dfrac{d\lambda_s}{\sqrt{\mathcal P(\lambda_s)}}=
\frac{2d\ell}{\prod_{i\neq s} (\lambda_s-\lambda_i)}.
$$
Thus, along a billiard trajectory, the differentials
$(-1)^{s-1}\dfrac{d\lambda_s}{\sqrt{\mathcal P(\lambda_s)}}$ stay
always positive, if we assume that the signs of the square roots
are chosen appropriately on each segment.

\smallskip

From these remarks and the discussion preceding this theorem, it
follows that the value of the integral
$\int\dfrac{\lambda_s^id\lambda_s}{\sqrt{\mathcal P(\lambda_s)}}$
between two consecutive common points of the trajectory and the
quadric $\mathcal Q_{\gamma_s'}$ (or $\mathcal Q_{\gamma_s''}$) is
equal to:
$$
2(-1)^{s-1} \int_{\gamma_s'}^{\gamma_s''}
\dfrac{\lambda_s^id\lambda_s}{+\sqrt{\mathcal P(\lambda_s)}}.
$$

Now, if $\mathbf p$ is a finite polygon representing a billiard
trajectory and having exactly $n_s$ points at $\mathcal
Q_{\gamma_s'}$ and $n_s$ at $\mathcal Q_{\gamma_s''}\ (1\le s\le
d)$, then
$$
\sum \int^{\mathbf p} \dfrac{\lambda_s^id\lambda_s}{\sqrt{\mathcal
P(\lambda_s)}}= 2\sum(-1)^{s-1}n_s
\int_{\gamma_s'}^{\gamma_s''}\dfrac{\lambda_s^id\lambda_s}{+\sqrt{\mathcal
P(\lambda_s)}}, \quad (1\le i\le d).
$$
Finally, the polygonal line is closed if and only if
$$
\sum (-1)^s
n_s\int_{\gamma_s'}^{\gamma_s''}\dfrac{\lambda_s^id\lambda_s}{\sqrt{\mathcal
P(\lambda_s)}}=0, \quad (1\le i\le d-1),
$$
which was needed. \end{proof}

\begin{example}
{\rm Consider two domains $\Omega'$ and $\Omega''$ in $\mathbb
R^3$. Let $\Omega'$ be bounded by the ellipsoid $\mathcal Q_0$ and
the two-folded hyperboloid $\mathcal Q_{\beta}$, $a_2<\beta<a_1$,
in such a way that $\Omega'$ is placed between the branches of
$\mathcal Q_{\beta}$. On the other hand, suppose $\Omega''$ is
bounded by $\mathcal Q_0$, the righthand branch of $\mathcal
Q_{\beta}$ (this one which is placed in the half-space $x_1>0$)
and the plane $x_1=0$. Elliptic coordinates of points inside both
$\Omega'$ and $\Omega''$ satisfy:
$$
0\le\lambda_3\le a_3,\ \ \beta\le\lambda_1\le a_1.
$$
Consider billiard trajectories within these two domains, with
caustics $\mathcal Q_{\mu_1}$ and $\mathcal Q_{\mu_2}$,
$a_3<\mu_1<a_2$, $a_2<\mu_2<a_1$. Since $\mu_2\le\beta$, the
segments $\Lambda_s$ ($s\in\{1,2,3\}$) of all possible values of
elliptic coordinates along a trajectory are, for both domains:
$$
\Lambda_1=[\beta,a_1],\ \Lambda_2=[\mu_1,a_2],\ \Lambda_3=[0,a_3].
$$
In $\Omega''$, existence of a periodic trajectory with caustics
$\mathcal Q_{\mu_1}$ and $\mathcal Q_{\mu_2}$, which becomes
closed after $n$ bounces at $\mathcal Q_0$ and $2m$ bounces at
$\mathcal Q_{\beta}$ is equivalent to the equality:
$$
n\bigl(\mathcal A(P_0) - \mathcal A(P_{a_3})\bigr)+
2m\bigl(\mathcal A(P_{\beta}) - \mathcal A(P_{\mu_1})\bigr)=0,
$$
on the Jacobian of the corresponding hyperelliptic curve. In
$\Omega'$, existence of a trajectory with same properties is
equivalent to:
$$
n\bigl(\mathcal A(P_0) - \mathcal A(P_{a_3})\bigr)+
2m\bigl(\mathcal A(P_{\beta}) - \mathcal A(P_{\mu_1})\bigr)=0 \ \
\text{and}\ \ n \ \ \text{is even}.
$$

\begin{figure}[h]
\centering
\begin{minipage}{0.44\textwidth}
\centering
\includegraphics[width=5cm,height=4cm]{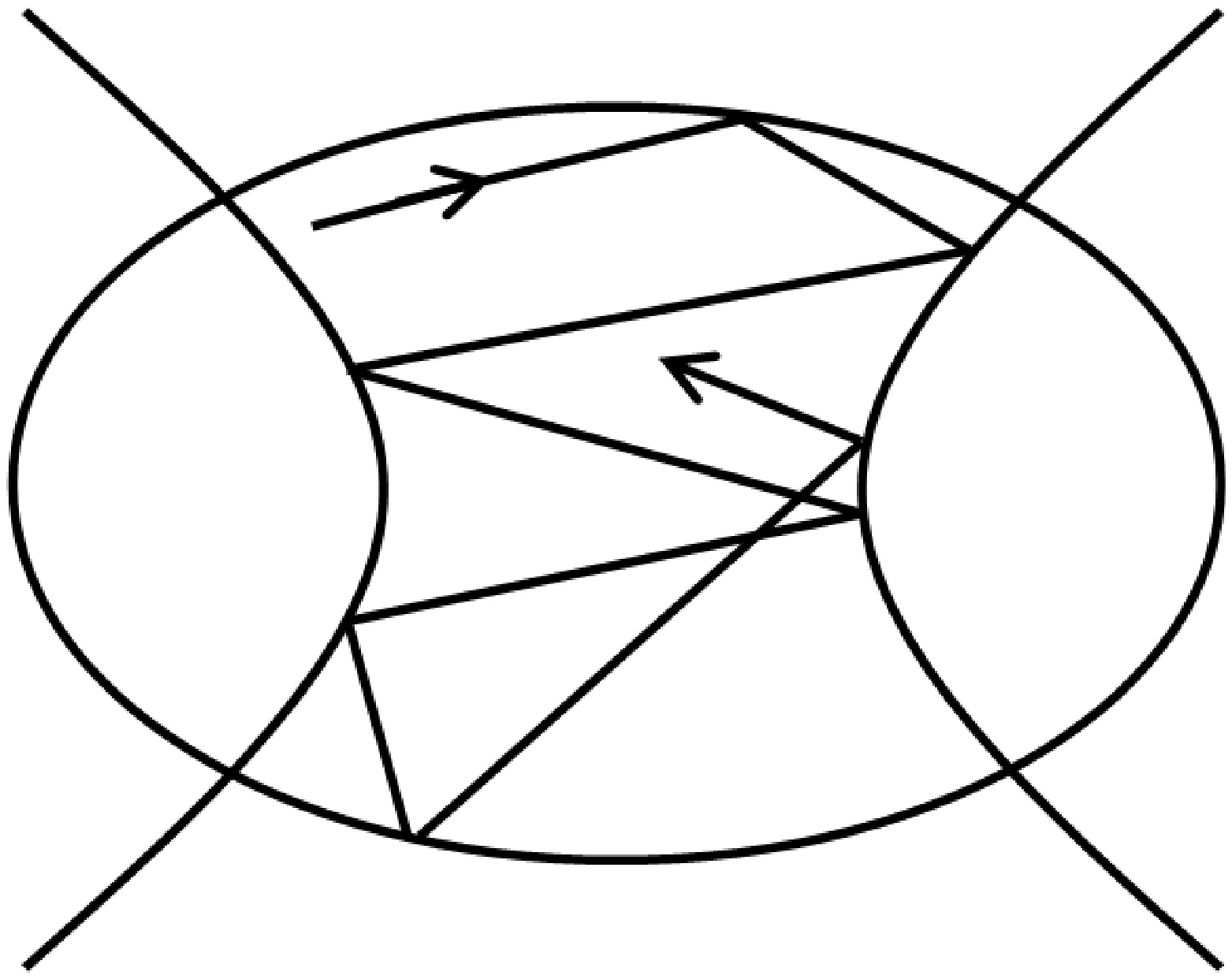}
\caption{A trajectory inside $\Omega'$}\label{fig:omega1}
\end{minipage}
\begin{minipage}{0.44\textwidth}
\centering
\includegraphics[width=5cm,height=4cm]{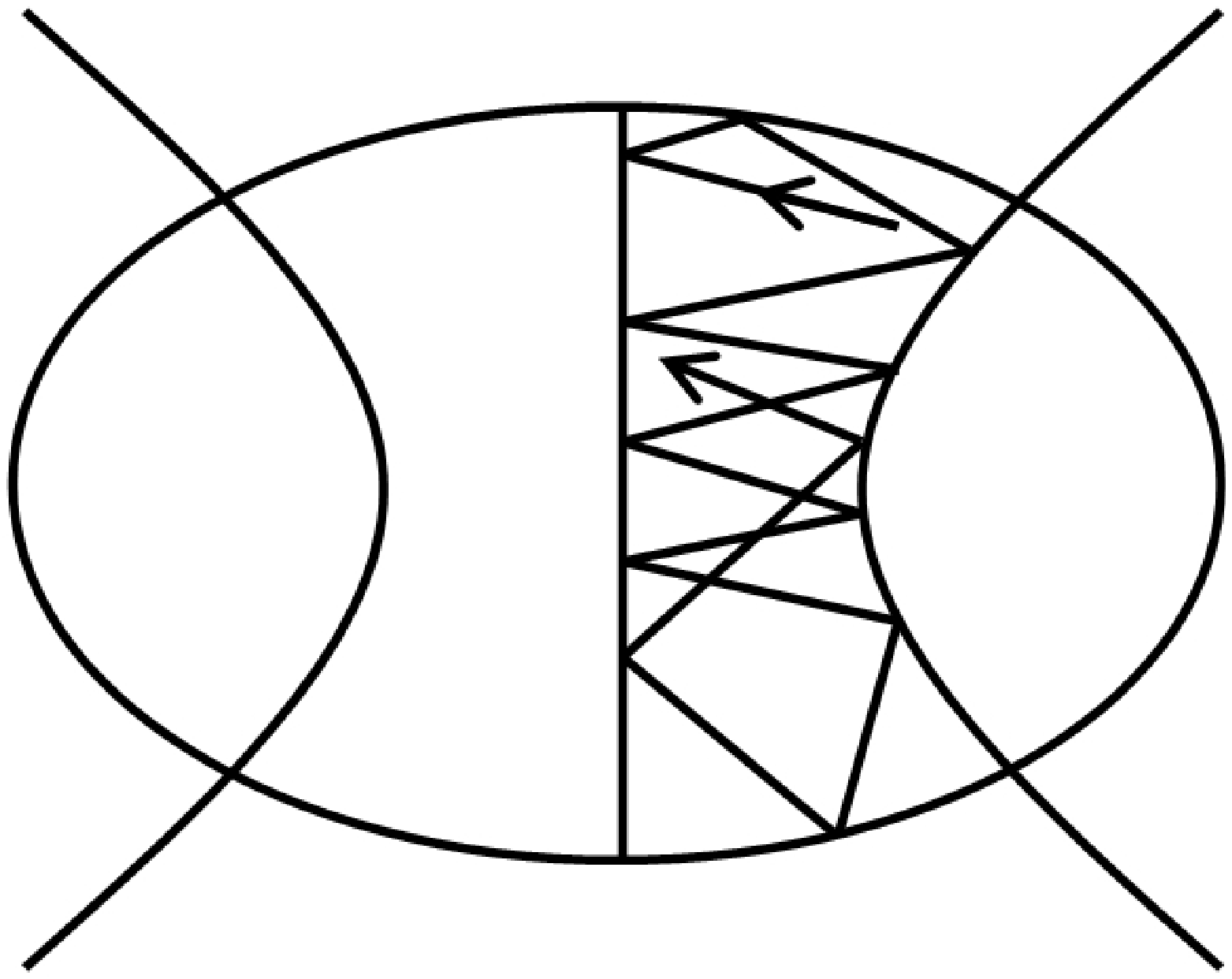}
\caption{A trajectory inside $\Omega''$}\label{fig:omega2}
\end{minipage}
\end{figure}
The fact that the second equality implies the first one is due to
the following geometrical fact: any billiard trajectory in
$\Omega'$ can be transformed to a trajectory in $\Omega''$
applying the symmetry with respect to the $x_2x_3$-plane to all
its points placed under this plane. Notice that this
correspondence of trajectories is 2 to 1 -- in such a way, a
generic billiard trajectory in $\Omega''$ corresponds to exactly
two trajectories in $\Omega'$. An example of such corresponding
billiard trajectories is shown on Figures \ref{fig:omega1} and
\ref{fig:omega2}. }
\end{example}

\subsection{Billiard Ordered Game}

Our  next step is to introduce a notion of bounces ``from
outside'' and ``from inside''. More precisely, let us consider an
ellipsoid $\mathcal Q_{\lambda}$ from the confocal family
(\ref{konfokalna.familija}) such that $\lambda \in (a_{s+1}, a_s)$
for some $s\in \{1,\dots, d\}$, where $a_{d+1}=-\infty$.

\smallskip

Observe that along a billiard ray which reflects at $\mathcal
Q_{\lambda}$, the elliptic coordinate $\lambda_i$ has a local
extremum at the point of reflection.

\begin{definition}
\begin{figure}[h]
\centering
\begin{minipage}{0.44\textwidth}
\centering
\includegraphics[width=5cm,height=4cm]{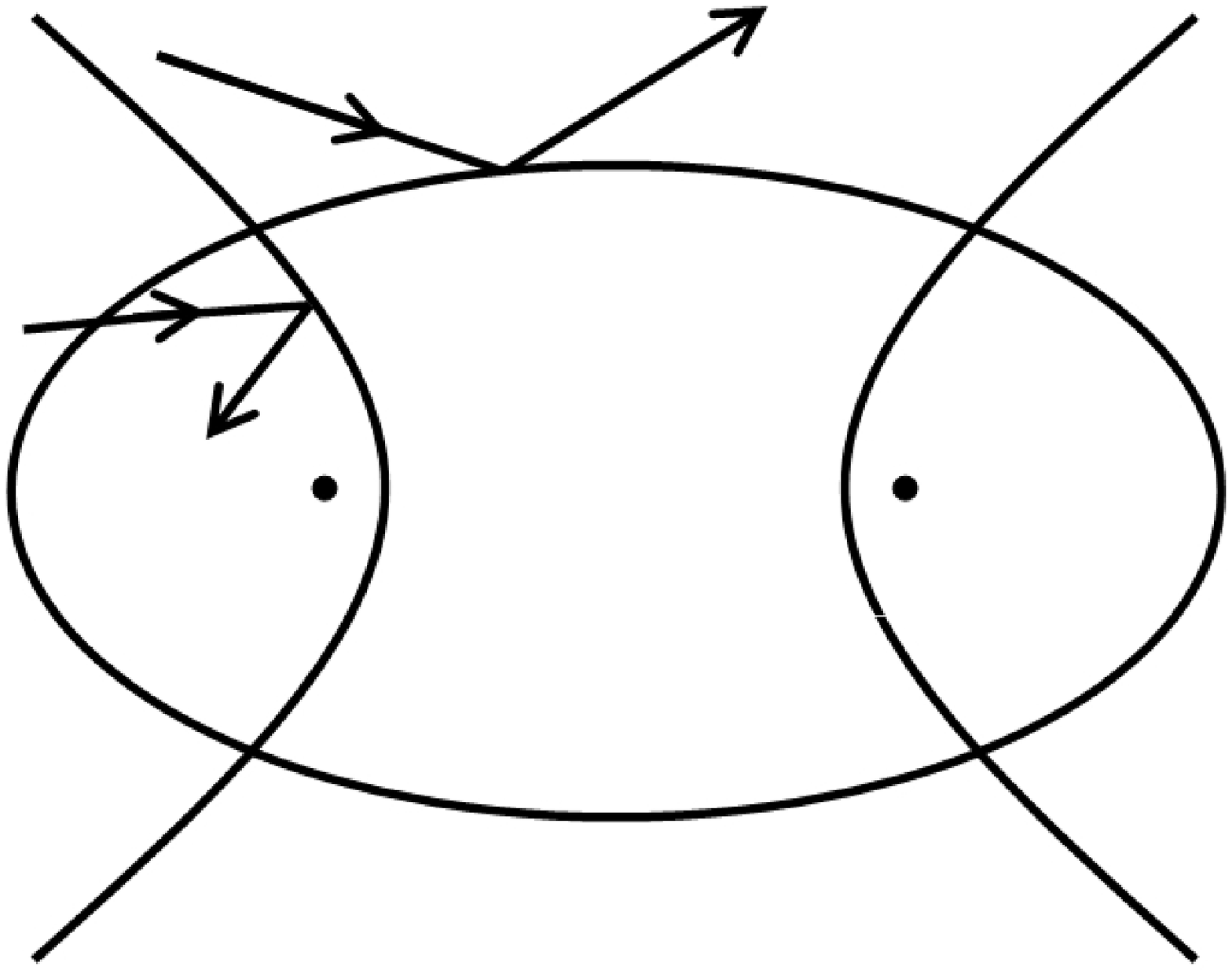}
\caption{Reflection from outside}\label{fig:odbijanje.spolja}
\end{minipage}
\begin{minipage}{0.44\textwidth}
\centering
\includegraphics[width=5cm,height=4cm]{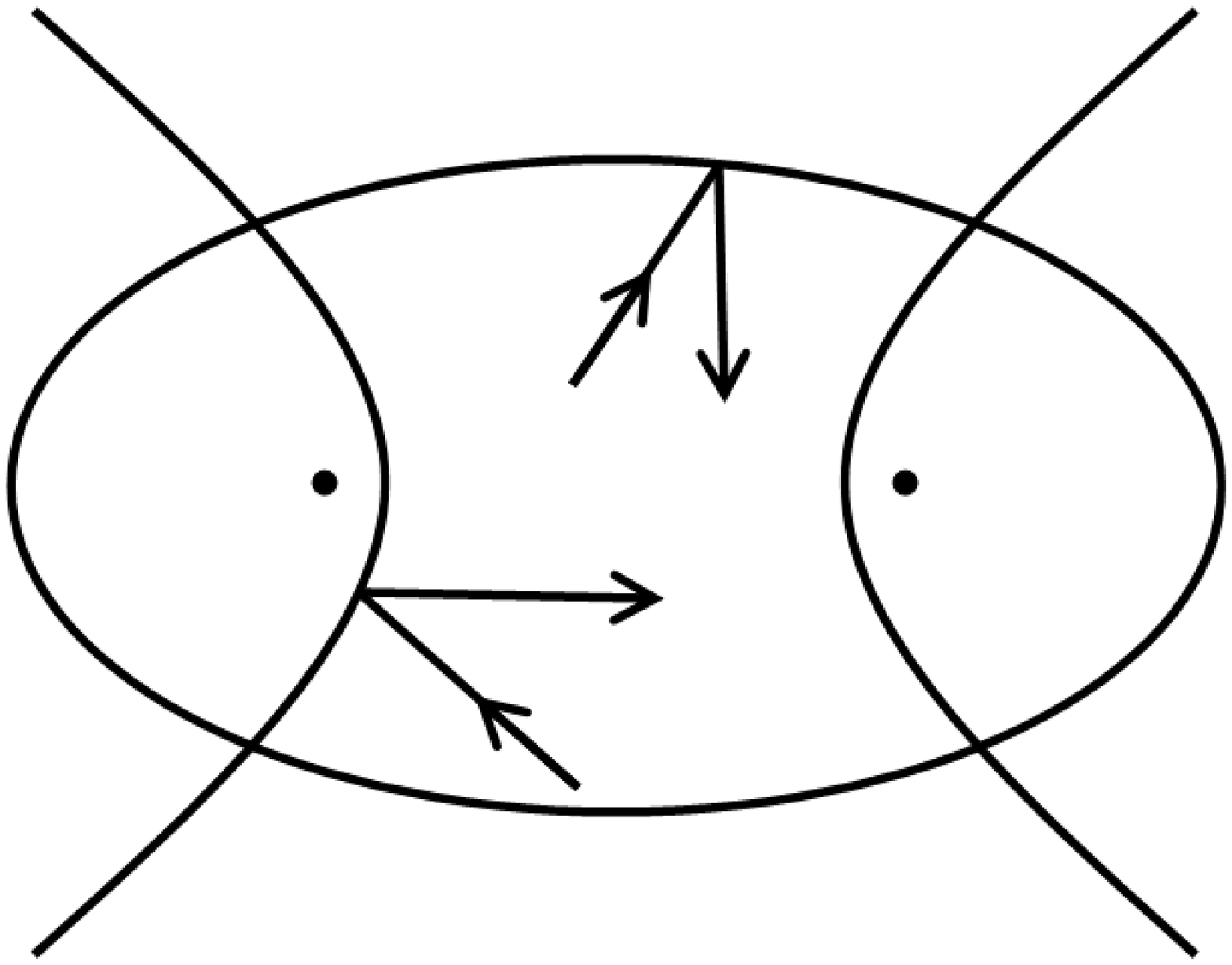}
\caption{Reflection from inside}\label{fig:odbijanje.iznutra}
\end{minipage}
\end{figure}
{\rm A ray reflects {\sl from outside} at the quadric $\mathcal
Q_{\lambda}$ if the reflection point is a local maximum of the
Jacobian coordinate $\lambda_s$, and it reflects {\sl from inside}
if the reflection point is a local minimum of the coordinate
$\lambda_s$.}
\end{definition}

On Figures \ref{fig:odbijanje.spolja} and
\ref{fig:odbijanje.iznutra} reflections from inside and outside an
ellipse and a hyperbola are sketched.

\smallskip

Let us remark that in the case when $\mathcal Q_{\lambda}$ is an
ellipsoid, the notions introduced in Definition 1 coincide with
the usual ones.

\smallskip

Assume now a $k$-tuple of confocal quadrics $\mathcal
Q_{\beta_1},\dots, \mathcal Q_{\beta_k}$ from the confocal pencil
(\ref{konfokalna.familija}) is given, and
$(i_1,\dots,i_k)\in\{-1,1\}^k$.

\begin{definition}{\rm
{\it The billiard ordered game} joined to quadrics $\mathcal
Q_{\beta_1},\dots,\mathcal Q_{\beta_k}$, with the {\it signature}
$(i_1,\dots,i_k)$ is the billiard system with trajectories having
bounces at $\mathcal Q_{\beta_1},\dots,\mathcal Q_{\beta_k}$
respectively, such that
$$
\displaylines{ \text {the reflection at}\; \mathcal Q_{\beta_s}\;
\text {is from inside if}\; i_s=+1;\cr
\text {the reflection at}\;
\mathcal Q_{\beta_s}\; \text{is from outside if}\; i_s=-1. }
$$
}
\end{definition}

Note that any trajectory of a billiard ordered game has $d-1$
caustics from the same family (\ref{konfokalna.familija}).

\smallskip

Suppose $\mathcal Q_{\beta_1}$, \dots, $\mathcal Q_{\beta_k}$ are
ellipsoids and consider a billiard ordered game with the signature
$(i_1,\dots,i_k)$. In order that trajectories of such a game stay
bounded, the following condition has to be satisfied:
$$
i_s=-1\ \Rightarrow\ i_{s+1}=i_{s-1}=1 \; \text{and}\;
\beta_{s+1}<\beta_s,\ \beta_{s-1}<\beta_s.
$$
(Here, we identify indices 0 and $k+1$ with $k$ and 1
respectively.)

\begin{example}
{\rm On Figure \ref{fig:billiard.game}, a trajectory corresponding
to the $7$-tuple
$$
(\mathcal Q_1, \mathcal Q_2, \mathcal Q_1,
\mathcal Q_3, \mathcal Q_2, \mathcal Q_3, \mathcal Q_1),
$$
with the signature $(1,-1,1,-1,1,1,1)$, is shown.
\begin{figure}[h]
\centering
\begin{minipage}{0.6\textwidth}
\centering
\includegraphics[width=5cm,height=4cm]{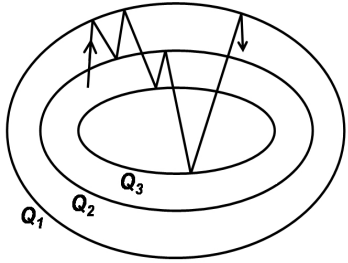}
\caption{Billiard ordered game}\label{fig:billiard.game}
\end{minipage}
\end{figure}
}
\end{example}

\begin{theorem}\label{uslov.igra} Given a billiard ordered game within $k$
ellipsoids $\mathcal Q_{\beta_1},\dots, \mathcal Q_{\beta_k}$ with
the signature $(i_1,\dots,i_k)$. Its trajectory with caustics
$\mathcal Q_{\alpha_1},\dots, \mathcal Q_{\alpha_{d-1}}$ is
$k$-periodic if and only if
$$
\sum_{s=1}^k i_s\bigl(\mathcal A(P_{\beta_s})-\mathcal
A(P_{\alpha})\bigr)
$$
is equal to a sum of several expressions of the form:
$\left(\mathcal A(P_{\alpha_p})-\mathcal
A(P_{\alpha_{p'}})\right)$ on the Jacobian of the curve $\Gamma\,
:\, y^2=\mathcal P(x),$ where $P_{\beta_s}=\left(\beta_s,+\sqrt
{\mathcal P(\beta_s)}\right)$, $\alpha =\min\{a_d,
\alpha_1,\dots,\alpha_{d-1}\}$ and $\mathcal Q_{\alpha_p}$,
$\mathcal Q_{\alpha_{p'}}$ are pairs of caustics of the same type.
\end{theorem}

When $\mathcal Q_{\beta_1}=\dotsb=\mathcal Q_{\beta_k}$ and
$i_1=\dotsb=i_k=1$ we obtain the Cayley-type condition for the
billiard motion inside an ellipsoid in $\mathbb R^d$.

\smallskip

We are going to treat in more detail the case of the billiard
motion between two ellipsoids.

\begin{proposition} The condition that there exists a closed
billiard trajectory between two ellipsoids $\mathcal Q_{\beta_1}$
and $\mathcal Q_{\beta_2}$, which bounces exactly $m$ times to
each of them, with caustics $\mathcal Q_{\alpha_1}, \dots,
\mathcal Q_{\alpha_{d-1}}$, is:
$$
\rank\left(\begin{array}{llll} f_1'(P_{\beta_2}) &
f_2'(P_{\beta_2}) & \dots &
f_{m-d+1}'(P_{\beta_2}) \\
f_1''(P_{\beta_2}) & f_2''(P_{\beta_2}) & \dots &
f_{m-d+1}''(P_{\beta_2}) \\
 & & \dots\\
 & & \dots\\
f_1^{(m-1)}(P_{\beta_2}) & f_2^{(m-1)}(P_{\beta_2}) & \dots &
f_{m-d+1}^{(m-1)}(P_{\beta_2})
\end{array}\right)<m-d+1.
$$
Here
$$
f_j=
\frac{y-B_0-B_1(x-\beta_1)-\dots-B_{d+j-2}(x-\beta_1)^{d+j-2}}{x^{d+j-1}},
\quad 1\le j\le m-d+1,
$$
and $y=B_0+B_1(x-\beta_1)+\dots$ is the Taylor expansion around
the point symmetric to $P_{\beta_1}$ with respect to the
hyperelliptic involution of the curve $\Gamma$. {\rm (All
notations are as in Theorem \ref{uslov.igra}.)}
\end{proposition}

\begin{example} {\rm Consider a billiard motion in the
three-dimensional space, with ellipsoids $\mathcal Q_0$ and
$\mathcal Q_{\gamma}$ as boundaries ($0<\gamma<a_3$) and caustics
$\mathcal Q_{\alpha_1}$ and $\mathcal Q_{\alpha_2}$. Such a motion
closes after 4 bounces from inside to $\mathcal Q_0$ and 4 bounces
from outside to $\mathcal Q_{\gamma}$ if and only if:
$$
\rank X< 2.
$$
The matrix $X$ is given by:
$$
\aligned
X_{11} &=-3C_0+C_1\gamma+3B_0+2B_1\gamma+B_2\gamma^2 \\
X_{12} &= -4C_0+C_1\gamma+4B_0+3B_1\gamma+2B_2\gamma^2+B_3\gamma^3\\
X_{21} &= 6C_0-3C_1\gamma+C_2\gamma^2-6B_0-3B_1\gamma-B_2\gamma^2\\
X_{22} &= 10C_0-4C_1\gamma-10B_0-6B_1\gamma-3B_2\gamma^2-B_3\gamma^3\\
X_{31} &= -10C_0+6C_1\gamma-3C_2\gamma^2+C_3\gamma^3+10B_0+4B_1\gamma+B_2\gamma^2\\
X_{32} &=
-20C_0-10C_1\gamma-4C_2\gamma^2+C_3\gamma^3+20B_0+10B_1\gamma+4B_2\gamma^2+B_3\gamma^3,
\endaligned
$$
and the expressions
$$
\aligned -\sqrt{(a_1-x)(a_2-x)(a_3-x)(\alpha_1-x)(\alpha_2-x)}
&= B_0+B_1x+B_2x^2+\dots,\\
+\sqrt{(a_1-x)(a_2-x)(a_3-x)(\alpha_1-x)(\alpha_2-x)} &=
C_0+C_1(x-\gamma)+\dots
\endaligned
$$
are Taylor expansions around points $x=0$ and $x=\gamma$
respectively.}
\end{example}

\begin{example}\label{ex:4iznutra}
 {\rm Using the same notations as in the
previous example, let us consider trajectories with 4 bounces from
inside to each of $\mathcal Q_0$ and $\mathcal Q_{\gamma}$, as
shown on Figure \ref{fig:4iznutra}.
\begin{figure}[h]
\centering
\begin{minipage}{0.6\textwidth}
\centering
\includegraphics[width=5cm,height=4cm]{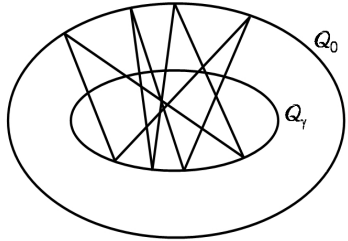}
\end{minipage}
\parbox{7cm}{\caption{A closed trajectory of the billiard ordered game with 8 alternate bounces from inside of two
ellipses}\label{fig:4iznutra}}
\end{figure}

The explicit condition for periodicity of such a trajectory is:
$$
\rank X<2,
$$
with
$$
\aligned
X_{11} &= -4C_0+C_1\gamma+3B_1\gamma+2B_2\gamma^2+B_3\gamma^3 \\
X_{12} &= -3C_0+C_1\gamma+3B_0+2B_1\gamma+B_2\gamma^2\\
X_{21} &= -6C_0+C_2\gamma^2+6B_0+6B_1\gamma+5B_2\gamma^2+3B_3\gamma^3\\
X_{22} &= -6C_0+C_1\gamma+C_2\gamma^2+6B_0+5B_1\gamma+3B_2\gamma^2\\
X_{31} &= -4C_0+C_3\gamma^3+4B_0+4B_1\gamma+4B_2\gamma^2+3B_3\gamma^3\\
X_{32} &=
-4C_0+C_2\gamma^2+C_3\gamma^3+4B_0+4B_1\gamma+3B_2\gamma^2.
\endaligned
$$}
\end{example}

\section{Elliptical Billiard as a Discrete Time System: $d$ --
Arbitrary, $k=1$}\label{discrete}

Another approach to the description of periodic billiard
trajectories is based on the technique of discrete Lax
representation.

\smallskip

In this section, first, we are going to list the main steps of
algebro-geometric integration of the elliptic billiard, following
\cite{MV}. Then, the connection between periodic billiard
trajectories and points of finite order on the corresponding
hyperelliptic curve will be established, and, using results from
Section \ref{hyper}, the Cayley-type conditions will be derived,
as they were obtained by the authors in \cite{DR1, DR2}. In
addition, here we provide a more detailed discussion concerning
trajectories with the period not greater than $d$ and the cases of
the singular isospectral curve.

\smallskip

Following \cite{MV}, the billiard system will be considered as a
system with the discrete time. Using its integration procedure,
the connection between periodic billiard trajectories and points
of finite order on the corresponding hyperelliptic curve will be
established.

\subsection{XYZ Model and Isospectral Curves}

\noindent{\bf Elliptical Billiard as a Mechanical System with the
Discrete Time.} Let the ellipsoid in ${\mathbb R}^d$ be given by
\begin{equation*}\label{elipsoid}
(Ax,x)=1.
\end{equation*}
We can assume that $A$ is a diagonal
matrix, with different eigenvalues. The billiard motion within the
ellipsoid is determined by the following equations:
$$ x_{k+1} - x_k = \mu_k y_{k+1} $$
$$ y_{k+1} - y_k = \nu_k A x_k,$$
where
$$ \mu_k = - \frac{2(A y_{k+1},x_k)}{(A y_{k+1},y_{k+1})}, \quad
   \nu_k = - \frac{2(A x_k, y_k)}{(A x_k, A x_k)}.
$$

Here, $x_k$ is a sequence of points of billiard bounces, while
$y_k = \dfrac{x_k - x_{k-1}}{|x_k - x_{k-1}|}$ are the momenta.

\

\noindent{\bf Connection between Billiard and XYZ Model.} To the
billiard system with the discrete time, Heisenberg $XYZ$ model can
be joined, in the way described by Veselov and Moser in \cite{MV}
and which is going to be presented here.

\smallskip

Consider the mapping $\varphi: (x,y) \mapsto (x',y')$ given by
$$
x_k' = J y_{k+1} = J(y_k + \nu_k A x_k), \quad y_k' = - J^{-1}x_k,
\quad J = A^{-\frac{1}{2}}.
$$
Notice that the dynamics of $\varphi$ contains the billiard
dynamics:
$$
x_k'' = J y_{k+1}' = -x_{k+1}, \quad y_k'' = - J^{-1}x_k' =
-y_{k+1},
$$
and define the sequence $(\bar x_k, \bar y_k)$:
$$
(\bar x_0, \bar y_0) := (x_0, y_0), \quad (\bar x_{k+1}, \bar
y_{k+1}) := \varphi(\bar x_k, \bar y_k),
$$
which obeys the following relations:
$$
\bar x_{k+1} = J \bar y_k + \nu_k J^{-1} \bar x_k, \quad \bar
y_{k+1} = - J^{-1} \bar x_k,
$$
where the parameter $\nu_k$ is such that $|\bar{y}_k|=1$,
$(A\bar{x}_k, \bar{x}_x)=1$. This can be rewritten in the
following way:
$$
\bar{x}_{k+1} + \bar x_{k-1} = \nu_k J^{-1} \bar x_k.
$$
Now, for the sequence $q_k := J^{-1} \bar x_k$, we have:
$$
 q_{k+1} + q_{k-1} = \nu_k J^{-1}q_k, \quad |q_k|=1.
$$
These equations represent the equations of the discrete Heisenberg
$XYZ$ system.

\begin{theorem} {\rm \cite{MV}} Let $(\bar x_k, \bar y_k)$ be the
sequence connected with elliptical billiard in the described way.
Then $q_k = J^{-1} \bar x_k$ is a solution of the discrete
Heisenberg system.

Conversely, if $q_k$ is a solution to the Heisenberg system, then
the sequence $x_k= (-1)^k J q_{2k}$ is a trajectory of the
discrete billiard within an ellipsoid.
\end{theorem}

\medskip

\noindent{\bf Integration of the Discrete Heisenberg XYZ System.}
Usual scheme of algebro-geometric integration contains the
following \cite{MV}. First, the sequence $L_k(\lambda)$ of matrix
polynomials has to be determined, together with a factorization
$$
   L(\lambda) = B(\lambda)C(\lambda) \mapsto C(\lambda)B(\lambda)
   = B'(\lambda)C'(\lambda) = L'(\lambda),
$$
such that the dynamics $L \mapsto L'$ corresponds to the dynamics
of the system $q_k$. For each problem, finding this sequence of
matrices requires a separate search and a mathematician with the
excellent intuition. All matrices $L_k$ are mutually similar, and
they determine the same {\sl isospectral curve}
$$
\Gamma \ : \ \det (L(\lambda) - \mu I) = 0.
$$
The factorization $L_k = B_k C_k$ gives splitting of spectrum of
$L_k$. Denote by $\psi_k$ the corresponding eigenvectors. Consider
these vectors as meromorphic functions on $\Gamma$ and denote
their pole divisors by $D_k$.

\smallskip

The sequence of divisors is linear on the Jacobian of the
isospectral curve, and this enables us to find, conversely,
eigenfunctions $\psi_k$, then matrices $L_k$, and, finally, the
sequence $(q_k)$.

\smallskip

Now, integration of the discrete $XYZ$ system by this method will
be shortly presented. Details of the procedure can be found in
\cite{MV}.

\smallskip

The equations of discrete $XYZ$ model are equivalent to the
isospectral deformation:
$$
L_{k+1}(\lambda) = A_k(\lambda) L_k(\lambda) A_k^{-1}(\lambda),
$$
where
$$
L_k(\lambda) = J^2 + \lambda q_{k-1} \wedge J q_k -\lambda ^2
q_{k-1} \otimes q_{k-1},
$$
$$
A_k(\lambda) = J - \lambda q_k \otimes q_{k-1}.
$$
The equation of the isospectral curve $\Gamma\, :\,
\det(L(\lambda) - \mu I) = 0$ can be written in the following
form:
\begin{equation}\label{kriva}
\nu^2 = \prod_{i=1}^{d-1} (\mu - \mu_i) \prod_{j=1}^d (\mu -
J_j^2),
\end{equation}
where $\nu = \lambda \prod_{i=1}^{d-1} (\mu
- \mu_i)$ and $\mu_1, \dots, \mu_{d-1}$ are zeroes of the
function:
$$
\phi_{\mu}(x,Jy) = \sum_{i=1}^d \frac{F_i (x,y)}{\mu - J_i^2},
$$
$$
F_i = x_i^2 + \sum_{j \neq i} \frac{ (x \wedge Jy)_{ij}^2 }{ J_i^2
- J_j^2 }, \quad  x=q_{k-1}, \quad y=q_k.
$$
It can be proved that $\mu_1,\dots,\mu_{d-1}$ are parameters of
the caustics corresponding to the billiard trajectory \cite{Mo}.
Another way for obtaining the same conclusion is to calculate them
directly by taking the first segment of the billiard trajectory to
be parallel to a coordinate axe.

\smallskip

If eigenvectors $\psi_k$ of matrices $L_k(\lambda)$ are known, it
is possible to determine uniquely members of the sequence $q_k$.
Let $D_k$ be the divisor of poles of function $\psi_k$ on curve
$\Gamma$. Then \cite{MV}:
$$
D_{k+1} = D_k + P_{\infty} -P_0,
$$
where $P_{\infty}$ is the point corresponding to the value $\mu =
\infty$ and $P_0$ to $\mu = 0$, $\lambda = (q_k, J^{-1}
q_{k+1})^{-1}$.

\subsection{Characterization of Periodical Billiard
Trajectories}

In the next lemmae, we establish a connection between periodic
billiard sequences $q_k$ and periodic divisors $D_k$.

\begin{lemma} {\rm \cite{DR1}} Sequence of divisors $D_k$ is
$n$-periodic if and only if the sequence $q_k$ is also periodic
with the period $n$ or $q_{k+n} = -q_k$ for all $k$.
\end{lemma}

\begin{proof} If, for all $k$, $q_{k+n}=q_k$, or
$q_{k+n}=-q_k$ for all $k$, then, obviously, $L_{k+n}=L_k$. Thus,
the sequence of eigenvectors $\psi_k$ is periodic with the period
$n$. It follows that the sequence of divisors is also periodic.

\smallskip

Suppose now that $D_{k+n}=D_k$ for all $k$. This implies that
$\psi_{k+n}=c_k \psi_k$. We have:
$$\psi_{k+1}=A_k(\lambda) \psi_k.$$
Let $\mu_1$ and $\mu_2$ be values of parameter $\mu$ which
correspond to the value $\lambda=1$ on the curve $\Gamma$, and
$$
\Psi_k=(\psi_k(1,\mu_1),\psi_k(1,\mu_2)).
$$
From $\psi_{k+1} = A_k(\lambda) \psi_k$ , we obtain $ A_k(1)=
\Psi_{k+1} \Psi_k^{-1}$. It follows that
$$
A_k(1)= \frac{c_{k+1}}{c_k} A_{k+n}(1).
$$
From the condition $ \det A_k= \det A_{k+1} $ for all $k$, we have
$c_k=c_{k+1}$. Thus, the sequence
$$ A_k(1)=J-q_k \otimes q_{k-1} $$
is $n$-periodic. From there,
$$
  q_{k+n} = \alpha_k q_k, \quad q_{k+n-1}= \frac{1}{\alpha_k} q_{k-1}.
$$
Since $|q_k|=1$, we have $\alpha_k=1$ or $\alpha_k=-1$, where all
$\alpha_k$ are equal to each other, which proves the assertion.
\end{proof}

\begin{lemma}\label{divizor} {\rm \cite{DR1}} The billiard is, up to the central
symmetry, periodic with the period $n$ if and only if the divisor
sequence $D_k$ joined to the corresponding Heisenberg $XYZ$ system
is also periodic, with the period $2n$.
\end{lemma}

\begin{proof} Let $x_{k+n} =\alpha x_k$ for all $k$,
$\alpha\in\{-1,1\}$. Join to a billiard trajectory $(x_k, y_k)$
the corresponding flow $(\bar x_k, \bar y_k)$. Since
$$
(\bar x_{2k},   \bar y_{2k})   = (-1)^k (x_k, y_k), \quad (\bar
x_{2k+1}, \bar y_{2k+1}) = \phi (\bar x_{2k}, \bar y_{2k}),
$$
where the mapping $\phi$ is linear, we obtain:
$$
\bar x_{k+2n} = \alpha(-1)^n \bar x_k.
$$

From there, immediately follows that $q_{k+2n} = \alpha(-1)^n
q_k$. According to Lemma 3, the divisor sequence $D_k$ is
$2n$-periodic. \end{proof}

Applying the previous lemma, we obtain the main statement of this
section:

\begin{theorem}\label{nas.uslov} {\rm \cite{DR2}} A condition on a billiard trajectory
inside ellipsoid $\mathcal Q_0$ in $\mathbb R^d$, with
non-degenerate caustics $\mathcal Q_{\mu_1},\dots,\mathcal
Q_{\mu_{d-1}}$, to be periodic, up to the central symmetry, with
the period $n\ge d$ is:
$$
\rank \left( \begin{array}{llll}
B_{n+1}  & B_n      &   \dots & B_{d+1}\\
B_{n+2}  & B_{n+1}  &   \dots & B_{d+2}\\
\hdotsfor 4                            \\
B_{2n-1} & B_{2n-2} &   \dots & B_{n+d-1}
\end{array} \right) < n-d+1,
$$
where
$$
\sqrt{(x-\mu_1)\cdots(x-\mu_{d-1})(x-a_1)\cdots(x-a_d)} =
     B_0 + B_1 x + B_2 x^2 + \dots .
$$
\end{theorem}

\begin{proof} The trajectory is periodic with period $n$
if, by Lemma \ref{divizor}, the corresponding divisor sequence on
the curve $\Gamma$ has the period $2n$, i.e.\
$2n(P_{\infty}-P_0)=0$ on $\mathcal J(\Gamma)$. Curve $\Gamma$ is
hyperelliptic with genus $g=d-1$. Taking $\mathcal A(P_{\infty})$
to be the neutral on $\mathcal J(\Gamma)$ we get the desired
result by applying Lemma \ref{nP=0}. \end{proof}

\smallskip

\noindent{\bf Cases of Singular Isospectral Curve.} When all $a_1,
\dots, a_d, \mu_1, \dots, \mu_{d-1}$ are mutually different, then
the isospectral curve has no singularities in the affine part.
However, singularities appear in the following three cases and
their combinations:

\smallskip

\noindent{\bf (i)} $a_i=\mu_j$ for some $i,j$. The isospectral
curve (\ref{kriva}) decomposes into a rational and a hyperelliptic
curve. Geometrically, this means that the caustic corresponding to
$\mu_i$ degenerates into the hyperplane $x_i=0$. The billiard
trajectory can be asymptotically tending to that hyperplane (and
therefore cannot be periodic), or completely placed in this
hyperplane. Therefore, closed trajectories appear when they are
placed in a coordinate hyperplane. Such a motion can be discussed
like in the case of dimension $d-1$.

\smallskip

\noindent{\bf (ii)} $a_i=a_j$ for some $i\neq j$. The boundary
$\mathcal Q_0$ is symmetric.

\smallskip

\noindent{\bf (iii)} $\mu_i=\mu_j$ for some $i\neq j$. The
billiard trajectory is placed on the corresponding confocal
quadric hyper-surface.

\smallskip

In the cases (ii) and (iii) the isospectral curve $\Gamma$ is a
hyperelliptic curve with singularities. In spite of their
different geometrical nature, they both need the same analysis of
the condition $2nP_0\sim 2nE$ for the singular curve
(\ref{kriva}).

\smallskip

Immediate consequence of Lemma \ref{sing.kriva} is that Theorem
\ref{nas.uslov} can be applied not only for the case of the
regular isospectral curve, but in the cases (ii) and (iii), too.
Therefore, the following interesting property holds.

\begin{theorem}\label{kratke.trajektorije}
If the billiard trajectory within an
ellipsoid in $d$-dimensional Eucledean space is periodic, up to
the central symmetry, with the period $n<d$, then it is placed in
one of the $n$-dimensional planes of symmetry of the ellipsoid.
\end{theorem}

\begin{proof} This follows immediately from Theorem
\ref{nas.uslov} and the fact that the section of a confocal family
of quadrics with a coordinate hyperplane is again a confocal
family. \end{proof}

Note that all trajectories periodic with period $n$ up to the
central symmetry, are closed after $2n$ bounces. A statement
sharper than Theorem \ref{kratke.trajektorije} for the
trajectories closed after $n$ bounces, where $n\le d$, can be
obtained in the elementary fashion:

\begin{proposition} If the billiard trajectory within an
ellipsoid in $d$-dimensional Eucledean space is periodic with
period $n\le d$, then it is placed in one of the
$(n-1)$-dimensional planes of symmetry of the ellipsoid.
\end{proposition}

\begin{proof} First, consider the case $n=d$. Let
$x_1\cdots x_d$ be a periodic trajectory, and $(N,x)=\alpha$ the
equation of the hyperplane spanned by its vertices. Here, $N$ is a
vector normal to the hyperplane and $\alpha$ is a constant. Since
all lines normal to the surface of the ellipsoid at the points of
reflection belong to this hyperplane, it follows that
$(AN,x_i)=(N,Ax_i)=0$. Thus, $(AN,x)=0$ is also an equation of the
hyperplane, so $\alpha=0$ and the vectors $N$, $AN$ are collinear.
From here, the claim follows immediately.

The case $n<d$ can be proved similarly, or applying Theorem
\ref{kratke.trajektorije} and the previous case of this
proposition. \end{proof}

This property can be seen easily for $d=3$.

\begin{example} {\rm Consider the billiard motion in an
ellipsoid in the $3$-dimensional space, with $\mu_1=\mu_2$, when
the segments of the trajectory are placed on generatrices of the
corresponding one-folded hyperboloid, confocal to the ellipsoid.
If there existed a periodic trajectory with period $n=d=3$, the
three bounces would have been coplanar, and the intersection of
that plane and the quadric would have consisted of three lines,
which is impossible. It is obvious that any periodic trajectory
with period $n=2$ is placed along one of the axes of the
ellipsoid. So, there is no periodic trajectories contained in a
confocal quadric surface, with period less or equal to $3$.}
\end{example}

\section{Periodic Trajectories of Billiards on Quadrics in
$\mathbf R^{\mathbf d}$}\label{on.quad}

In \cite{CS}, the billiard systems on a quadric $\mathcal Q_0$ in
$\mathbb R^d$:
$$
\frac {x_1^2}{a_1}+\dots + \frac {x_d^2}{a_d}=1,\quad
a_1>\dotsb>a_d,
$$
are defined as limits of corresponding billiards within $\mathcal
Q_0$, when one of the caustics tends to $\mathcal Q_0$. The
boundary of such a billiard consists of the intersection of
$\mathcal Q_0$ with certain confocal quadrics $\mathcal
Q_{\beta_1}, \dots, \mathcal Q_{\beta_k}$ of the family
(\ref{konfokalna.familija}). Between the bounces, the billiard
trajectories follow geodesics of the surface of $\mathcal Q_0$,
while obeying the reflection law at the points of the boundary. A
tangent line issued from any point of a given trajectory touches,
besides $\mathcal Q_0$, also $d-2$ confocal quadrics $\mathcal
Q_{\alpha_1}$, \dots, $\mathcal Q_{\alpha_{d-2}}$ from the family
(\ref{konfokalna.familija}). These $d-2$ quadrics are fixed for
each trajectory, and we shall refer to them as caustics.

\smallskip

The question of description of periodic trajectories of these
systems was formulated as an open problem by Abenda and Fedorov
\cite{Ab}.

\smallskip

Like in Section \ref{k.quad}, we will consider two cases: the
billiard inside a domain $\Omega\subset\mathcal Q_0$ bounded by
several confocal quadrics, and the billiard ordered game within
several confocal quadrics of the same type.

\smallskip

For the first case, the domain $\Omega$ is given by:
$$
\beta_1'\le\lambda_1\le\beta_1'',\quad \dots,\quad
\beta_{d-1}'\le\lambda_{d-1}\le\beta_{d-1}'', \quad (\beta_d=0),
$$
where $\beta_s',\beta_s''\in[a_{s+1},a_s]$ for $1\le s\le d-1$.
Like in Section \ref{k.quad}, $\Lambda_s$, $(1\le s\le d-1)$, is
defined as the set of all values taken by the coordinate
$\lambda_s$ on the given trajectory, and it can be shown in the
same way that
$$
\Lambda_s=\{\, \lambda\in[\beta_s',\beta_s'']\, :\, \mathcal
P_1(\lambda)\ge0\,\},
$$
with
$$
\mathcal
P_1(x)=-x(a_1-x)\cdots(a_d-x)(\alpha_1-x)\cdots(\alpha_{d-2}-x),
$$
where $\alpha_1$, \dots, $\alpha_{d-2}$ are parameters of caustics
of the trajectory.

Denote $[\gamma_s',\gamma_s'']:=\Lambda_s$.

\smallskip

Before formulating the theorem, let us define the following
projection of the Abel-Jacobi map on the curve
$$
\Gamma_1 \ :\ y^2=\mathcal P_1(x),
$$
by:
$$
\bar{\mathcal{A}}(P)= \left(\begin{array}{c}
0 \\
\int_0^P \dfrac{x dx}y  \\
\int_0^P \dfrac{x^2 dx}y   \\
\dots \\
\int_0^P \dfrac{x^{d-2} dx}y
\end{array}\right).
$$

\begin{theorem}\label{omega.ellipsoid}
A trajectory of the billiard system constrained
on the ellipsoid $\mathcal Q_0$ within $\Omega$, with caustics
$\mathcal Q_{\alpha_1}$, \dots, $\mathcal Q_{\alpha_{d-2}}$, is
periodic with exactly $n_s$ bounces at each of quadrics $\mathcal
Q_{\gamma_s'}$, $\mathcal Q_{\gamma_s''}$, $(1\le s\le d-1)$, if
and only if
$$
\sum_{s=1}^{d-1} n_s\bigl(\bar{\mathcal A}(P_{\gamma_s'})
-\bar{\mathcal A}(P_{\gamma_s''})\bigr)=0.
$$
Here $P_{\gamma_s'}$, $P_{\gamma_s''}$ are the points on
$\Gamma_1$ with coordinates $P_{\gamma_s'}=\left(\gamma_s',
(-1)^s\sqrt {\mathcal P_1(\gamma_s')}\right)$,
$P_{\gamma_s''}=\left(\gamma_s'', (-1)^s\sqrt {\mathcal
P_1(\gamma_s'')}\right)$.
\end{theorem}

\begin{proof} By \cite{Jac}, the system of differential
equations of a geodesic line on $\mathcal Q_0$ with the caustics
$\mathcal Q_{\alpha_1}$, \dots, $\mathcal Q_{\alpha_{d-2}}$ is:
$$
\sum_{s=1}^{d-1}\frac{\lambda_s d\lambda_s}{\sqrt{\mathcal
P_1(\lambda_s)}}=0, \quad \sum_{s=1}^{d-1}\frac{\lambda_s^2
d\lambda_s}{\sqrt{\mathcal P_1(\lambda_s)}}=0, \quad \dots, \quad
\sum_{s=1}^{d-1}\frac{\lambda_s^{d-2} d\lambda_s}{\sqrt{\mathcal
P_1(\lambda_s)}}=0,
$$
with the square root $\sqrt{\mathcal P(\lambda_s)}$ taken with the
same sign in all of the expressions, for any fixed $s$. Also,
$$
\sum_{s=1}^{d-1}\frac{\lambda_s^{d-1} d\lambda_s}{\sqrt{\mathcal
P_1(\lambda_s)}}=2d\ell,
$$
where $d\ell$ is the length element.

Now, the rest of the demonstration is completely parallel to the
proof of Theorem \ref{uslov.k}. \end{proof}

In the same way as in Section \ref{k.quad}, a billiard ordered
game constrained on the ellipsoid $\mathcal Q_0$ within given
quadrics $\mathcal Q_{\beta_1}, \dots, \mathcal Q_{\beta_k}$ of
the same type can be defined. The only difference is that now the
signature $\sigma=(i_1,\dots,i_k)$ can be given arbitrarily, since
trajectories are bounded, lying on the compact hypersurface
$\mathcal Q_0$. Denote by $\mathcal Q_{\alpha_1}$, \dots,
$\mathcal Q_{\alpha_{d-2}}$ the caustics of a given trajectory of
the game. Since quadrics $\mathcal Q_{\beta_1}, \dots, \mathcal
Q_{\beta_k}$ are all of the same type, there exist $\mu',\mu''$ in
the set $S=\{a_1,\dots,a_d,\alpha_1,\dots,\alpha_{d-2}\}$ such
that $\beta_1,\dots,\beta_k\in[\mu',\mu'']$ and $(\mu',\mu'')\cap
S$ is empty.

\smallskip

Associate to the game the following divisors on the curve
$\Gamma_1$:
$$
\mathcal D_s= \begin{cases}
P_{\mu''}, & \text{if}\; i_s=i_{s+1}=1\\
0, &
\begin{array}{l} \text{if}\; i_s=-i_{s+1}=1, \beta_s<\beta_{s+1}\\
  \text{or}\; i_s=-i_{s+1}=-1, \beta_s>\beta_{s+1}
\end{array}
  \\
P_{\mu''}-P_{\mu'}, & \text{if}\; i_s=-i_{s+1}=1, \beta_s>\beta_{s+1}\\
P_{\mu'}-P_{\mu''}, & \text{if}\; i_s=-i_{s+1}=-1, \beta_s<\beta_{s+1}\\
P_{\mu'}, & \text{if}\; i_s=i_{s+1}=-1,
\end{cases}
$$
where $P_{\mu'}$ and $P_{\mu''}$ are the branching points with
coordinates $(\mu',0)$ and $(\mu'',0)$ respectively.

\begin{theorem}\label{game.ellipsoid}
Given a billiard ordered game constrained on $\mathcal Q_0$ within
quadrics $\mathcal Q_{\beta_1}$, \dots, $\mathcal Q_{\beta_k}$
with signature $\sigma=(i_1,\dots,i_k)$. Its trajectory with
caustics $\mathcal Q_{\alpha_1}$, \dots, $\mathcal
Q_{\alpha_{d-2}}$ is $k$-periodic if and only if
$$
\sum_{s=1}^k i_s\bigl(\bar{\mathcal A}(P_{\beta_s})-\bar{\mathcal
A}(\mathcal D_s)\bigr)
$$
is equal to a sum of several expressions of the form
$\bar{\mathcal A}(P_{\alpha_p})-\bar{\mathcal A}(P_{\alpha_{p'}})$
on the Jacobian of the curve $ \Gamma_1 : y^2=\mathcal P_1(x), $
where $P_{\beta_s}=\left(\beta_s,+\sqrt {\mathcal
P_1(\beta_s)}\right)$ and $Q_{\alpha_p}$, $Q_{\alpha_{p'}}$ are
pairs of caustics of the same type.
\end{theorem}

\begin{example}
{\rm Consider the case $d=3$ and a billiard system constrained on
the ellipsoid $\mathcal Q_0$ with the boundary $\mathcal
Q_{\gamma}$ and caustic $\mathcal Q_{\alpha}$,
$a_3<\gamma<\alpha<a_2$. A sufficient condition for a
corresponding billiard trajectory to be $n$-periodic is:
$$
n\bigl({\mathcal A}(P_{\gamma})-{\mathcal
A}(P_{\alpha})\bigr)=0,
$$
or, equivalently, in Cayley-type form:
$$
\rank\left(\begin{array}{llll}
C_{p+1} & C_{p+2} & \dots & C_{2p-2}\\
C_{p+2} & C_{p+3} & \dots & C_{2p-1}\\
\dots\\
C_{2p} & C_{2p+1} & \dots & C_{3p-3}
\end{array}\right)<p-2,
\qquad
n=2p,
$$
and
$$
\rank\left(\begin{array}{llll}
C_{p+1} & C_{p+2} & \dots & C_{2p-1}\\
C_{p+2} & C_{p+3} & \dots & C_{2p}\\
\dots\\
C_{2p} & C_{2p+1} & \dots & C_{3p-2}
\end{array}\right)<p-1,
\qquad
n=2p+1,
$$
where
$$
y=C_0 + C_1\left(\tilde x-{\frac1{\alpha}-\gamma}\right)+C_2\left(\tilde x-{\frac1{\alpha}-\gamma}\right)^2+\dots,
$$
is the Taylor expansion around the point $P_{\gamma}$,
with $\tilde x=\dfrac1{\alpha-x}$. (See also \cite{DR3}.)

More precisely, this condition will be satisfied if and only if the trajectory is $n$-periodic and its length $L$ with respect to the parameter $s$:
$$
ds=\lambda_1\lambda_2\cdots\lambda_nd\ell
$$
is such that the vector
$$
\left(
\begin{array}{c}
L/2\\
0\\
\vdots\\
0
\end{array}\right)
+n\bar{\mathcal A}(P_{\gamma})- n\bar{\mathcal A}(P_{\alpha})
$$
belongs to the period-lattice of the Jacobian of the corresponding hyper-elliptic curve.
}
\end{example}

\begin{example}\label{ex:jura}
{\rm In a recent preprint \cite{Ab}, another interesting class of
periodical billiard trajectories on 2-dimensional ellipsoid,
associated with closed geodesics, is described. Such geodesics are
closely connected with 3-elliptic KdV solutions, which
historically goes back to Hermite and Halphen \cite{HH}.

\smallskip

\noindent
The corresponding genus 2 isospectral curve is:
\begin{equation}\label{kriva.G}
\mathrm G\ :\ y^2=-\frac14(4x^3-9g_2x-27g_3)(x^2-3g_2).
\end{equation}
This curve is $3:1$ tangetial covering of two elliptic curves:
$$
\displaylines{ \mathrm E_1\ :\ y_1^2=4x_1^3 - g_2 x_1 - g_3,\cr
\mathrm E_2\ :\
y_2^2=x_2^3-\frac{27}{16}(g_2^3+9g_3^2)x_2-\frac{243}{32}(3g_3^2-g_3g_2^3).
}
$$
The corresponding conditions for closed billiard trajectories are
obtained in \cite{Ab}, by use of Lemma 3 from \cite{DR2}, which
corresponds to Lemma \ref{nP=0} of the present paper. For the
exact formulae and details of calculations, see \cite{Ab}. }
\end{example}

Note that in the same algebro-geometric situation as in the
previous example, with the application of Theorems
\ref{omega.ellipsoid} and \ref{game.ellipsoid}, it is possible to
describe a wider class of closed billiard trajectories, without
the assumption that their segments belong to closed geodesics.
Namely, an important property of the covering $\mathrm G\to\mathrm
E_1$ from Example \ref{ex:jura} is that the holomorphic
differential $\frac{xdx}y$ on $\mathrm G$ reduces to
$-\frac23\frac{dx_1}{y_1}$ on $\mathrm E_1$. Thus, whenever the
isospectral curve can be represented in the form (\ref{kriva.G}),
it is possible to obtain conditions of Cayley's type by the
application of Theorem \ref{omega.ellipsoid}, Theorem
\ref{game.ellipsoid} and Lemma \ref{nP=0}.

\smallskip

This is going to be illustrated in the next example.

\begin{example}{\rm
Consider the billiard motion on quadric $\mathcal Q_0$ in $\mathbb
R^3$, with $a_1=a\sqrt3$, $a_2=\frac32a$, $a_3=-a\sqrt3$, with the
caustic $\mathcal Q_{\alpha}$, $\alpha=-\frac32a$, $a>0$. The
domain $\Omega$ is bounded by confocal ellipsoid $\mathcal
Q_{\beta}$, $\beta<-a\sqrt3$. By Theorem \ref{omega.ellipsoid},
the condition for such a billiard to be $n$-periodic is:
\begin{equation}\label{eq:uslov.G}
n\bigl(\bar{\mathcal A}(P_{\beta}) -\bar{\mathcal
A}(P_{-a\sqrt3})\bigr)=0
\end{equation}
on the Jacobian of the curve
$$
y^2=-x\left(x^2-\frac94a^2\right)(x^2-3a^2).
$$
This curve is of the form \ref{kriva.G}, with $g_2=a^2$, $g_3=0$.

\smallskip

The covering $\mathrm G\to\mathrm E_1$ is given by:
$$
x_1=-\frac19\frac{x^3}{x^2-3a^2},\quad
y_1=\frac2{27}\frac{x^3-9a^2x}{(x^2-3a^2)^2}.
$$
The condition (\ref{eq:uslov.G}) reduces to the following
relation:
$$
n\bigl({\mathcal A}(Q_{\bar\beta}) - {\mathcal
A}(Q_{\infty})\bigr)=0
$$
on the Jacobian of the curve:
$$
\mathrm E_1\ :\ y_1^2=4x_1^3-a^2x_1,
$$
with the point $Q_{\bar\beta}\in\mathrm E_1$ having the
coordinates $(z_1=\bar\beta,\ w_1=4\bar\beta^3-a^2\bar\beta)$,
$\bar\beta=-\dfrac19\dfrac{\beta^3}{\beta^2-3a^2}$ and
$Q_{\infty}$ being the infinite point on $\mathrm E_1$.

\smallskip

By Lemma \ref{nP=0}, the desired condition becomes:
$$
\aligned
\left | \begin{array}{llll}
     B_3     & B_4     & \dots & B_{m+1} \\
     B_4     & B_5     & \dots & B_{m+2} \\
      & & \dots                          \\
     B_{m+1} & B_{m+2} & \dots & B_{2m-1}
     \end{array} \right |\, & =\, 0,
\quad \text{for}\ n=2m\\
\left | \begin{array}{llll}
     B_2     & B_3     & \dots & B_{m+1} \\
     B_3     & B_4     & \dots & B_{m+2} \\
      & & \dots                         \\
     B_{m+1} & B_{m+2} & \dots & B_{2m}
     \end{array} \right |\, & =\, 0,
\quad \text{for}\ n=2m+1,
\endaligned
$$
where $ \sqrt{4x_1^3-a^2x_1}=B_0 + B_1(x_1-\bar\beta) +
B_2(x_1-\bar\beta)^2+\dots$ is the Taylor expansion around the
point $Q_{\bar\beta}$.}
\end{example}

We will finish this section by analysis of the behaviour of geodesic lines on quadrics  after the reflection of a confocal quadric.

\smallskip

Denote by $\mathbf g$ a geodesic line on the quadric $\mathcal Q_0$ in $\mathbb R^d$, and by $\Omega$ a domain on $\mathcal Q_0$ bounded by a single confocal quadric $\mathcal Q_{\beta}$. We will suppose that the set $\mathbf g\cap\mathcal Q_{\beta}\neq\emptyset$ and that the geodesic $\mathbf g$ intersects $\mathcal Q_{\beta}$ transversally. Under these assumptions, the number of their intersection points will be finite if and only if the line $\mathbf g$ is closed. These points are naturally divided into two sets -- one set, denote it by $S_1$, contains those points where $\mathbf g$ enters into $\Omega$, while $S_2$ contains the points where the geodesic leaves the domain.

\smallskip

Consider reflections on $\mathcal Q_{\beta}$ according to the billiard rule in each point of $S_2$. Applying these reflections to $\mathbf g$, we obtain a family of geodesic segments on $\mathcal Q_0$. It appears that all these segments belong to one single geodesic line, see Figure \ref{fig:bilijar.geodezijska}.

\begin{proposition}\label{prop:bilijar.geodezijska}
Let $\mathbf{g'}$ be a geodesic line which contains a point $P\in S_2$, such that $\mathbf g$ and $\mathbf{g'}$ satisfy the reflection law in $P$ at $\mathcal Q_{\beta}$. Then $\mathbf{g'}$ contains all points of $S_2$ and the two lines $\mathbf g$, $\mathbf{g'}$ satisfy the reflection law at $\mathcal Q_{\beta}$ in each of these points.

If $\mathbf g$ is closed, then $\mathbf{g'}$ is also closed and they have the same number of intersection points with $\mathcal Q_{\beta}$.
\begin{figure}[h]
\centering
\begin{minipage}[b]{0.6\textwidth}
\centering
\includegraphics[width=5cm,height=4cm]{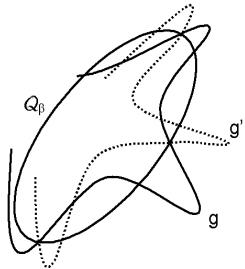}
\caption{Reflection of a geodesic line}\label{fig:bilijar.geodezijska}
\end{minipage}
\end{figure}
\end{proposition}

\begin{proof}
The reflection from inside at $\mathcal Q_{\beta}$ corresponds to the shift by the vector
$\bar{\mathcal A}(P_{\beta}) - \bar{\mathcal A}(\tau P_{\beta})$ on the Jacobian of the curve $\Gamma_1$, where $\tau:\Gamma_1\to\Gamma_1$ is the hyper-elliptic involution: $(x,y)\mapsto(x,-y)$. Thus, the shift does not depend on the choice of the intersection point and it maps one geodesic into another one. This proof is essentially the same as the proof of Lemma 5.1 in \cite{Ab}, which covers the special case of a closed geodesic line on the two-dimensional ellipsoid.
\end{proof}

Observe that $\mathbf{g'}$ not necessarily contains the points of $S_1$, see Figure \ref{fig:bilijar.geodezijska}.

\smallskip

Denote by $T_{\beta}$ the above mapping $\mathbf g\mapsto \mathbf{g'}$ in the set of geodesics on $\mathcal Q_0$ (on geodesic lines that are tangent to $\mathcal Q_{\beta}$ or do not intersect it at all, we define $T_{\beta}$ to be the identity).
As a consequence of Proposition \ref{prop:bilijar.geodezijska}, we obtain a new porism of Poncelet type.

\begin{corollary}
Suppose that $T_{\beta}^n(\mathbf g)=\mathbf g$. Then $T_{\beta}^n$ is the identity on the class of all geodesics sharing the same caustics with $\mathbf g$.

Moreover, if the geodesic line $\mathbf g$ is closed, then any billiard trajectory in $\Omega$ with the initial segment lying on $\mathbf g$, will be periodic.
\end{corollary}

\section{Virtual Billiard Trajectories}\label{virtual}

Apart from reflections from inside and outside of a quadric, as we
defined in Section 4, which correspond to the real motion of the
billiard particle, it is of interest to consider {\it virtual
reflections}. These reflections were considered by Darboux in
\cite{Dar3}. The aim of this section is to prove and generalize a
property of virtual reflections formulated by Darboux in the
three-dimensional case in the footnote \cite{Dar3} on p.\ 320-321:

\smallskip

{\it ``\dots Il importe de remarquer: le th\'eor\`eme donn\'e dans
le texte suppose essentiellement que les c\^ot\'es du polygone
soient form\'es par les parties {\bf r\'eelles} et non {\bf
virtuelles} du rayon r\'efl\'echi. Il existe des polygones
ferm\'es d'une tout autre nature. \'Etant donn\'es, par exemple,
deux ellipso{\"\i}des homofocaux $(E_0)$, $(E_1)$, si, par une
droite quelconque, on leur m\`ene des plans tangents, on aura
quatre points de contact $a_0$, $b_0$ sur $(E_0)$, $a_1$, $b_1$
sur $(E_1)$. Le quadrilat\`ere $a_0 a_1 b_0 b_1$ sera tel que les
bissectrices des angles $a_1$, $b_1$ soient les normales de
$(E_1)$, et les bissectrices des angles $a_0$, $b_0$ les normales
de $(E_0)$, mais il ne constituera pas une route {\bf r\'eelle}
pour un rayon lumineux; deux de ses c\^ot\'es seront form\'es par
les parties virtuelles des rayons r\'efl\'echis. De tels polygones
m\'eriteraient aussi d'\^etre \'etudi\'es, leur th\'eorie offre
les rapports les plus \'etroits avec celle de l'addition des
fonctions hyperelliptiques et de certaines surfaces alg\'ebriques
\dots ''}

\smallskip

More formally, we can define the {\it virtual reflection} at the
quadric $\mathcal Q$ as a map of a ray $\ell$ with the endpoint
$P_0$ ($P_0\in\mathcal Q$) to the ray complementary to the one
obtained from $\ell$ by the real reflection from $\mathcal Q$ at
the point $P_0$.

\smallskip

Let us remark that, in the case of real reflections, exactly one
elliptic coordinate, the one corresponding to the quadric
$\mathcal Q$, has a local extreme value at the point of
reflection. On the other hand, on a virtual reflected ray, this
coordinate is the only one not having a local extreme value. In
the 2-dimensional case, the virtual reflection can easily be
described as the real reflection from the other confocal conic
passing through the point $P_0$. In higher dimensional cases, the
virtual reflection can be regarded as the real reflection of the
line normal to $\mathcal Q$ at $P_0$.

\smallskip

From now on, we consider the $n$-dimensional case.

\smallskip

Let points  $X_1, X_2$; $Y_1, Y_2$ belong to quadrics $\mathcal
Q_1$, $\mathcal Q_2$ of a given confocal system.

\begin{definition}{\rm
We will say that the quadruple of points $X_1, X_2, Y_1, Y_2$
constitutes a {\it virtual reflection configuration} if pairs of
lines $X_1 Y_1$, $X_1 Y_2$; $X_2 Y_1$, $X_2 Y_2$; $X_1 Y_1$, $X_2
Y_1$; $X_1 Y_2$, $X_2 Y_2$ satisfy the reflection law at points
$X_1$, $X_2$ of $\mathcal Q_1$ and $Y_1$, $Y_2$ of $\mathcal Q_2$
respectively.
\begin{figure}[h]
\centering
\begin{minipage}[b]{0.6\textwidth}
\centering
\includegraphics[width=5cm,height=4cm]{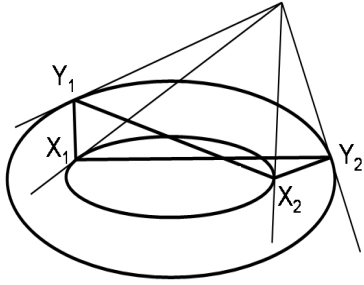}
\caption{Virtual reflection configuration}
\end{minipage}\label{fig:virtual_reflection}
\end{figure}
}
\end{definition}

Now, the Darboux's statement can be generalized and proved as
follows:

\begin{theorem}\label{th:virt.refl} Let $\mathcal Q_{\lambda_1}$, $\mathcal
Q_{\lambda_2}$ be confocal quadrics, $X_1$, $X_2$ points on
$\mathcal Q_{\lambda_1}$ and $Y_1$, $Y_2$ on $\mathcal
Q_{\lambda_2}$. If the tangent hyperplanes at these points to the
quadrics belong to a pencil, then $X_1, X_2, Y_1, Y_2$ constitute
a virtual reflection configuration.

Furthermore, if $\mathcal Q_{\lambda_1}$ and $\mathcal
Q_{\lambda_2}$ are ellipsoids and $\lambda_1>\lambda_2$, then the
sides of the quadrilateral $X_1Y_1X_2Y_2$ obey the real reflection
from $\mathcal Q_{\lambda_2}$ and the virtual reflection from
$\mathcal Q_{\lambda_1}$.
\end{theorem}

\begin{proof} Denote by $\xi_1$, $\xi_2$ and $\eta_1$,
$\eta_2$ the tangent hyperplanes to $\mathcal Q_{\lambda_1}$ at
$X_1$, $X_2$ and to $\mathcal Q_{\lambda_2}$ at $Y_1$, $Y_2$,
respectively. All these hyperplanes belong to a pencil, thus their
poles with respect to any quadric will be collinear --
particularly, the pole $P$ of $\xi_1$ lies on line $Y_1Y_2$. If
$Q=Y_1Y_2\cap\xi_1$, then pairs $P$, $Q$ and $Y_1$, $Y_2$ are
harmonically conjugate. It follows that the lines $X_1Y_1$,
$X_1Y_2$ obey the reflection law from $\xi_1$. We prove similarly
that all other adjacent sides of the quadrilateral $X_1Y_1X_2Y_2$
obey the law of reflection on the corresponding quadrics.

\smallskip

Now, suppose that $\mathcal Q_{\lambda_1}$,
 $\mathcal Q_{\lambda_2}$
are ellipsoids, and $\lambda_1>\lambda_2$. This means that
 $\mathcal Q_{\lambda_1}$
is placed inside $\mathcal Q_{\lambda_2}$, thus the whole
quadrilateral $X_1Y_1X_2Y_2$ is inside $\mathcal Q_{\lambda_2}$.
This means that the reflections in points $Y_1$, $Y_2$ are real
reflections from inside on $\mathcal Q_{\lambda_2}$. Besides, the
ellipsoid $\mathcal Q_{\lambda_1}$ is completely placed inside the
dihedron with the sides $\eta_1$, $\eta_2$. This ellipsoid is also
inside the dihedron $\angle(\xi_1, \xi_2)$. Since planes $\eta_1$
i $\eta_2$ are outside $\angle(\xi_1, \xi_2)$, it follows that
points $Y_1$, $Y_2$ are also outside this dihedron. Thus, points
$Y_1$, $Y_2$ are placed at the different sides of each of the
planes $\xi_1$, $\xi_2$, and reflections of
 $\mathcal Q_{\lambda_1}$ are virtual.
\end{proof}

We are going to conclude this section with the statement converse
to the previous theorem.

\begin{proposition} Let pairs of points $X_1$, $X_2$ and
$Y_1$, $Y_2$ belong to confocal ellipsoids $\mathcal Q_1$ and
$\mathcal Q_2$, and let $\alpha_1$, $\alpha_2$, $\beta_1$,
$\beta_2$ be the corresponding tangent planes. If a quadruple
$X_1, X_2, Y_1, Y_2$ is a virtual reflection configuration, then
planes $\alpha_1$, $\alpha_2$, $\beta_1$, $\beta_2$ belong to a
pencil.
\end{proposition}

\begin{proof} Consider the pencil determined by $\alpha_1$
and $\beta_1$. Let $\alpha_2'$, $\beta_2'$ be planes of this
pencil, tangent respectively to $\mathcal Q_1$, $\mathcal Q_2$ at
points $X_2'$, $Y_2'$, and distinct from $\alpha_1$ and $\beta_1$.
By Theorem \ref{th:virt.refl}, the quadruple $X_1, X_2', Y_1,
Y_2'$ is a virtual reflection configuration. Moreover, if denote
by $\lambda_1$, $\lambda_2$ parameters of $\mathcal Q_1$,
$\mathcal Q_2$ and assume $\lambda_1>\lambda_2$, then the sides of
the quadrangle obey the reflection law at points $Y_1, Y_2'$ and
the virtual reflection at $X_1, X_2'$. Since the ray obtained from
$X_1Y_1$ by the virtual reflection of $\mathcal Q_1$ at $X_1$, has
only one intersection with $\mathcal Q_2$, we have $Y_2=Y_2'$.
Points $X_2$ and $X_2'$ coincide, being the intersection of rays
obtained from $Y_1X_1$ and $Y_2X_1$, by the reflection at the
quadric $\mathcal Q_2$. Now, the four tangent planes are all in
one pencil. \end{proof}

\section{On Generalization of Lebesgue's Proof of Cayley's
Condition}\label{gen.lebeg}

In this section, the analysis of possibility of generalization of
the inspiring Le\-besgue's procedure to higher dimensional cases
will be given.

\smallskip

A higher dimensional analogue of the crucial lemma from
\cite{Leb}, which is Lemma \ref{lebeg.lema} of this article, is
the following:

\begin{lemma}\label{mali.uopsten.lebeg}
Let $\mathcal Q_1$, $\mathcal Q_2$ be quadrics of a confocal
system and let lines $\ell_1$, $\ell_2$ satisfy the reflection law
at point $X_1$ of $\mathcal Q_1$ and $\ell_2$, $\ell_3$ at $Y_2$
of $\mathcal Q_2$. Then line $\ell_1$ meets $\mathcal Q_2$ at
point $Y_1$ and $\ell_3$ meets $\mathcal Q_1$ at point $X_2$ such
that pairs of lines $\ell_1$, $Y_1X_2$ and $Y_1X_2$,$\ell_3$
satisfy the reflection law at points $Y_1$, $X_2$ of quadrics
$\mathcal Q_2$, $\mathcal Q_1$ respectively. Moreover, tangent
planes at $X_1$, $X_2$, $Y_1$, $Y_2$ of these two quadrics are in
the same pencil.
\end{lemma}

This statement can be proved by the direct application of Theorem
\ref{th:virt.refl} on virtual reflections. Nevertheless, there is
no complete analogy between Lemma \ref{mali.uopsten.lebeg} and the
corresponding assertion in plane. Lines $\ell_1$, $\ell_3$ and
$\ell_2$, $Y_1 Y_2$ are generically skew. Hence we do not have the
third pair of planes tangent to the quadric, containing
intersection points of these two pairs of lines.

\smallskip

Nevertheless, a complete generalization of the Basic Lemma, can be
formulated as follows:

\begin{theorem}\label{uopsten.lebeg}
Let $\mathcal F$ be a dual pencil of quadrics in the
three-dimensional space.

For a given quadric $\Gamma_0\in\mathcal F$, there exist
quadruples $\alpha$, $\beta$, $\gamma$, $\delta$ of planes tangent
to $\Gamma_0$, and quadrics $\Gamma_1,\Gamma_2,
\Gamma_3\in\mathcal F$ touching the pairs of intersecting lines
$\alpha\beta$ and $\gamma\delta$, $\alpha\gamma$ and
$\beta\delta$, $\alpha\delta$ and $\beta\gamma$ respectively, with
the tangent planes to $\Gamma_1$, $\Gamma_2$, $\Gamma_3$ at points
of tangency with the lines, all being in one pencil $\Delta$.
Moreover, the six intersecting lines are in one bundle.

Every such a configuration of planes $\alpha$, $\beta$, $\gamma$,
$\delta$ and quadrics $\Gamma_0$, $\Gamma_1$, $\Gamma_2$,
$\Gamma_3$ is determined by two of the intersecting lines, and the
tangent planes to $\Gamma_1$, $\Gamma_2$, or $\Gamma_3$
corresponding to these two lines.
\end{theorem}

\begin{proof} Suppose first that two adjacent lines
$\alpha\beta$ and $\alpha\gamma$ are given. There exist two
quadrics in $\mathcal F$ touching $\alpha\beta$ --- let $\Gamma_1$
be the one tangent to the given plane $\mu\supset\alpha\beta$.
$\Gamma_2$ denotes the quadric that touches $\alpha\gamma$ and the
given plane $\pi\supset\alpha\gamma$. The pencil $\Delta$ is
determined by its planes $\mu$ and $\pi$. All three lines
$\alpha\beta$, $\alpha\gamma$, $\Delta$ are in one bundle
$\mathcal B$, see Figure \ref{fig:pramen}.

\begin{figure}[h]
\centering
\begin{minipage}{0.6\textwidth}
\centering
\includegraphics[width=7cm,height=4cm]{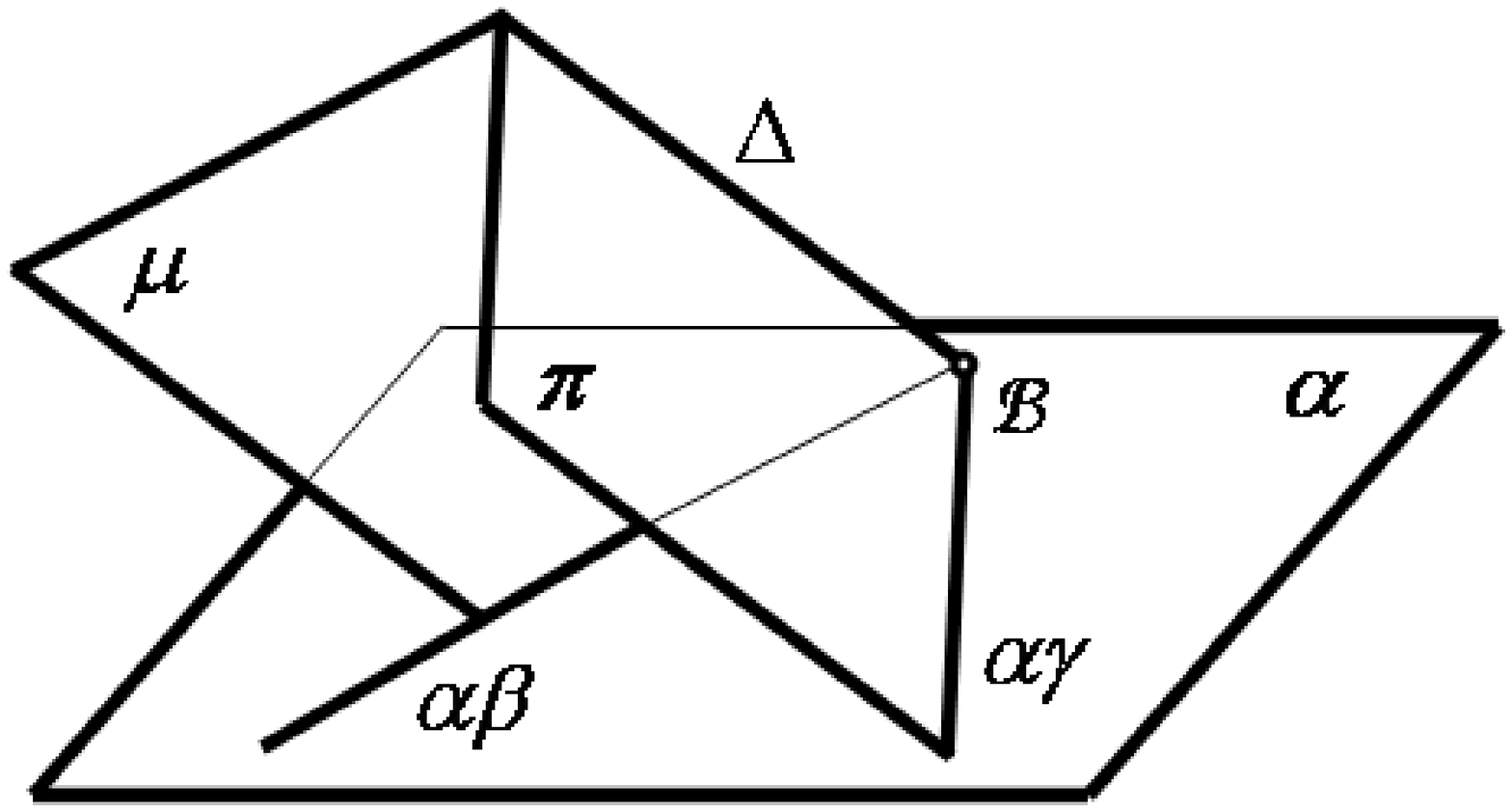}
\caption{Theorem \ref{uopsten.lebeg}}\label{fig:pramen}
\end{minipage}
\end{figure}

\smallskip

Note that lines $\alpha\beta$ and $\alpha\gamma$ determine plane
$\alpha$ and that $\alpha$ touches a unique quadric $\Gamma_0$
from $\mathcal F$. Thus, $\beta$ and $\gamma$ are determined as
tangent planes to $\Gamma_0$, containing lines $\alpha\beta$ and
$\alpha\gamma$ respectively and being different from $\alpha$.

\smallskip

Let $\nu$ be the plane from $\Delta$, other than $\mu$, that is
touching $\Gamma_1$. We are going to prove that the point of
tangency $\nu$ with $\Gamma_1$ belongs to $\gamma$.

\smallskip

Denote by $\Phi$ the dual pencil determined by quadric $\Gamma_1$
and the degenerate quadric which consists of lines $\alpha\beta$
and $\gamma\nu$. Since $\gamma\nu\in\mathcal B$, these two lines
are coplanar.

\smallskip

Dual pencils $\mathcal F$ and $\Phi$ determine the same involution
of the pencil $\alpha\gamma$, because they both determine the pair
$\alpha,\gamma$ and the quadric $\Gamma_1$ is the common for both
pencils. Since quadric $\Gamma_2\in\mathcal F$ determines the pair
$\pi,\pi$ of  coinciding planes, a quadric $\mathcal Q$
determining the same pair has to exist in $\Phi$. This quadric, as
well as all other quadrics in $\Phi$, touches $\nu$ and $\mu$.
Since $\nu$, $\mu$, $\pi$ all belong to the pencil $\Delta$,
$\mathcal Q$ is degenerate and contains $\Delta$. Since any
quadric of $\Phi$ touches $\mu$, $\pi$ at points of lines
$\alpha\beta$, $\alpha\gamma$ respectively, the other component of
$\mathcal Q$ also has to be $\Delta$, thus $\mathcal Q$ is the
double $\Delta$.

\smallskip

It follows that all quadrics of $\Phi$, and particularly
$\Gamma_1$, are tangent to $\nu$ at a point of $\gamma\nu$.

\smallskip

Similarly, if $\kappa\neq\pi$ is the other plane in $\Delta$
tangent to $\Gamma_2$, the touching point belongs to $\beta$ and
to $\delta$, the plane, other than $\gamma$, tangent to $\Gamma_0$
and containing the line $\gamma\nu$.

\smallskip

Now, let us note that $\mathcal F$ and $\Phi$ determine the same
involution on pencil $\alpha\delta$, because they both determine
pair $\alpha,\delta$ and $\Gamma_1$ belongs to both pencils. Thus,
the common plane $\rho$ of pencils $\Delta$ and $\alpha\delta$ is
tangent to a quadric of $\mathcal F$ at a point of $\alpha\delta$.
Denote this quadric by $\Gamma_3$. Similarly as before, we can
prove that $\Gamma_3$ is touching the line $\beta\gamma$ and the
corresponding tangent plane $\sigma$ is common to pencils $\Delta$
and $\beta\gamma$.

\smallskip

Now, suppose that two non-adjacent lines $\alpha\beta$,
$\gamma\delta$, both tangent to a quadric $\Gamma_1\in\mathcal F$,
with the corresponding tangent planes $\mu$, $\nu$, are given.
Similarly as above, we can prove that the plane $\rho$ is tangent
to a quadric from $\mathcal F$ at a point of $\alpha\delta$. So,
we can consider the configuration as determined by adjacent lines
$\alpha\beta$, $\alpha\delta$ and planes $\mu$, $\rho$. In this
way, we reduced it to the previous case.
\end{proof}

For the conclusion, we will generalize the notion of Cayley's
cubic.

\begin{definition}{\rm
The {\it generalized Cayley curve} is the variety of hyperplanes
tangent to quadrics of a given confocal family in $\mathbb CP^d$
at the points of a given line $\ell$.
}
\end{definition}

This curve is naturally embedded in the dual space $\mathbb
CP^{d\,*}$.

\smallskip

\begin{figure}[h]
\centering
\begin{minipage}{0.6\textwidth}
\centering
\includegraphics[width=7cm,height=4cm]{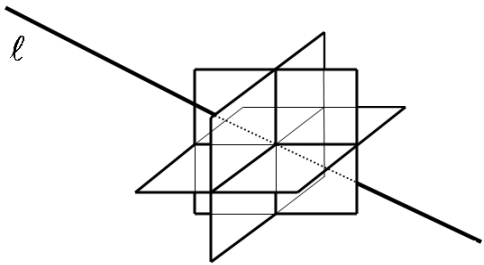}
\caption{Three points of the generalized Cayley curve in dimension
3}\label{fig:gen.cayley.curve}
\end{minipage}
\end{figure}

On Figure \ref{fig:gen.cayley.curve}, we see the planes which
correspond to one point of the line $\ell$ in the 3-dimensional
space.

\begin{proposition} The generalized Cayley curve in
$\mathbb C^d$, for $d\ge 3$ is a hyperelliptic curve of genus
$g=d-1$. Its natural realization in $\mathbb C^{d\,*}$ is of
degree $2d-1$.
\end{proposition}

\begin{proof}
Let us consider the projection from the generalized Cayley's curve
to the line of the parameters of the confocal family. Since $\ell$
intersects each quadric twice, this is a two-folded covering, with
branching points corresponding to quadrics touching $\ell$ and the
degenerate ones. Since there is $(d-1)+(d+1)$ such quadrics, we
obtain the genus directly from Riemann-Hurwitz formula.

\smallskip

Its degree is equal to the number of intersection points with a
hyperplane in $\mathbb C^{d\,*}$. Such a hyperplane is a bundle of
hyperplanes containing one point in $\mathbb C^d$. Take $P\in\ell$
to be this point. Since there are $d$ quadrics from the confocal
family containing $P$ and $d-1$ tangent to $\ell$, the assertion
follows.
\end{proof}

Let us note that this curve is isomorphic to the Veselov-Moser
isospectral curve (\ref{kriva}). Also, in the 3-dimensional case,
it is isomorphic to the Jacobi hyperelliptic curve, which was used
by Darboux considering the generalization of Poncelet theorem.

\smallskip

Further development of these ideas will be presented in the
separate publication \cite{DR4}.

\section*{Appendix 1\newline Integrable Potential Perturbations of the
Elliptical Billiard}
\addtocontents{toc}{\numberline{}{\bf Appendix 1\newline}}
\addtocontents{toc}{\numberline{}{\bf Integrable Potential
Perturbations of the Elliptical Billiard\hfill40}\newline}

The equation
\begin{equation}\label{BerDar}
\lambda V_{xy}+3\left( yV_x-xV_y\right) +(y^2-x^2)V_{xy}+xy\left(
V_{xx}- V_{yy}\right)=0,
\end{equation}
 is a special case of the
Bertrand-Darboux equation \cite{B, Dar2, Wh}, which represents the
necessary and sufficient condition for a natural mechanical system
with two degrees of freedom
$$
H=\frac{1}2(p^2_x + p^2_y) + V(x,y)
$$
to be separable in elliptical coordinates or some of their
degenerations.

\smallskip

Solutions of Equation (\ref{BerDar}) in the form of Laurent
polynomials in $x,y$ were described in \cite{Drag1}. Such
solutions are in \cite{Drag2} naturally related to the well-known
hypergeometric functions of Appell. This relation automatically
gives a wider class of solutions of Equation (\ref{BerDar}) -- new
potentials are obtained for non-integer parameters, giving a huge
family of integrable billiards within an ellipse with potentials.
Similar formulae for potential perturbations for the Jacobi
problem for geodesics on an ellipsoid from \cite{Drag1, Drag3} are
given. They show the existence of a connection between
separability of classical systems on one hand, and the theory of
hypergeometric functions on the other one, which is still not
completely understood. Basic references for the Appell functions
are \cite{Ap, AK, VK}.

\smallskip

The function $F_4$ is one of the four hypergeometric functions in
two variables, which are introduced by Appell \cite{Ap, AK}:
$$
F_4(a,b,c,d;x,y)=\sum \frac{(a)_{m+n} (b)_{m+n}}{(c)_m (d)_n}
\frac {x^m}{m!} \frac{y^n}{n!},
$$
where $(a)_n$ is the standard Pochhammer symbol:
$$
\aligned
(a)_n&=\frac {\Gamma (a+n)}{\Gamma (a)}=a(a+1)\dots (a+n-1),\\
(a)_0&=1.
\endaligned
$$
(For example $ m!=(1)_m.$)

\smallskip

The series is convergent for $\sqrt x + \sqrt y \le 1$. The
functions $F_4$ can be analytically continued to the solutions of
the equations:
$$
\aligned x(1-x)\frac{\partial^2 F}{\partial x^2} &- y^2
\frac{\partial^2 F}{\partial y^2}-2xy \frac{\partial^2 F}{\partial
x\partial y}+\\
 &+[c-(a+b+1)x]\frac{\partial F}{\partial x} - (a+b+1)y\frac{\partial F}{\partial y}-abF=0,\\
y(1-y)\frac{\partial^2 F}{\partial y^2} &- x^2 \frac {\partial^2
F}{\partial x^2}-2xy \frac {\partial^2F}{\partial x\partial
y}+\\
&+[c'-(a+b+1)y]\frac{\partial F}{\partial y}-(a+b+1)x
\frac{\partial F}{\partial x}-ab F=0 ,
\endaligned
$$

\subsection*{A1.1 Potential Perturbations of a Billiard inside
an Ellipse}

A billiard system which describes a particle moving freely within
the ellipse
$$
\frac{x^2}A + \frac{y^2}B = 1
$$
is completely integrable and it has an additional integral
$$
K_1=\frac{\dot x^2}A+\frac{\dot y^2}B-\frac{(\dot x y -\dot y
x)^2}{AB}.
$$
We are interested now in potential perturbations $V=V(x,y)$ such
that the perturbed system has an integral $\tilde K_1$ of the form
$$
 \tilde K_1=K_1 + k_1(x,y),
$$
where $k_1=k_1(x,y)$ depends only on coordinates. This specific
condition leads to Equation (\ref{BerDar}) on $V$ with
$\lambda=A-B$.

\smallskip

In \cite{Drag1, Drag3} the Laurent polynomial solutions of
Equation (\ref{BerDar}) were given. Denoting
\begin{equation}\label{Vgama}
V_{\gamma}=\tilde y ^{-\gamma}\bigl( (1-\gamma)\,\tilde x\, F_4(1,
2-\gamma, 2, 1-\gamma, \tilde x, \tilde y)+1\bigr),
\end{equation}
where $\tilde x=x^2/\lambda$, $\tilde y =-y^2/\lambda$, the more
general result was obtained in \cite{Drag2}:

\begin{theorem}\label{th:Vgama1} Every function $V_{\gamma}$ given with
{\rm (\ref{Vgama})} and $\gamma \in \mathbb C$ is a solution of
Equation {\rm (\ref{BerDar})}.
\end{theorem}

This theorem gives new potentials for non-integer $\gamma$. For
integer $\gamma$, one obtains the Laurent solutions.

\

\noindent{\bf Mechanical Interpretation.} With $\gamma\in\mathbb R^-$
and the coefficient multiplying $V_{\gamma}$ positive, we have a
potential barrier along $x$-axis. We can consider billiard motion
in the upper half-plane. Then we can assume that  a cut is done
along negative part of $y$-axis, in order to get a unique-valued
real function as a potential.

\smallskip

Solutions of Equation (\ref{BerDar}) are also connected with
interesting geometric subjects.

\subsection*{A1.2 The Jacobi Problem for Geodesics on an
Ellipsoid}

The Jacobi problem for the geodesics on an ellipsoid
$$
\frac{x^2}A + \frac{y^2}B + \frac{z^2}C=1
$$
has an additional integral
$$
K_1=\left(\frac{x^2}{A^2}+\frac{y^2}{B^2}+\frac{z^2}{C^2}\right)
\left(\frac{\dot x^2}A+\frac{\dot y^2}B+\frac{\dot z^2}C\right).
$$
Potential perturbations $V=V(x,y,z)$, such that perturbed systems
have integrals of the form
$$
\tilde K_1 = K_1 + k(x,y,z),
$$
satisfy the following system:
\begin{equation}\label{pde.sistem}
\aligned
\left(\frac{x^2}{A^2}+\frac{y^2}{B^2}+\frac{z^2}{C^2}\right)&
V_{xy}\frac{A-B}{AB} - 3\frac y{B^2}\frac{V_x}A +
3\frac{x}{A^2}\frac{V_y}B +\\
+\left(\frac{x^2}{A^3} - \frac {y^2}{B^3}\right)& V_{xy}
+\frac{xy}{AB} \left(\frac{V_{yy}}A - \frac {V_{xx}}B\right)
+\frac{zx}{CA^2}V_{zy}-\frac {zy}{CB^2}V_{zx}=0\\
\\
\left(\frac{x^2}{A^2}+\frac{y^2}{B^2}+\frac{z^2}{C^2}\right)&
V_{yz}\frac{B-C}{BC} -
3\frac{z}{C^2}\frac{V_y}B+3\frac{y}{B^2}\frac {V_z}C +\\
+\left(\frac {y^2}{B^3}-\frac {z^2}{C^3}\right)&V_{yz}
+\frac{yz}{BC}\left(\frac{V_{zz}}B-\frac {V_{yy}}C\right) +
\frac{xy}{AB^2}V_{xz}-\frac {xz}{AC^2}V_{xy}=0\\
\\
\left(\frac{x^2}{A^2} + \frac{y^2}{B^2} + \frac{z^2}{C^2}\right)&
V_{zx}\frac{C-A}{AC}-3\frac x{A^2}\frac{V_z}C+3\frac z{C^2}\frac
{V_x}A+\\
+\left(\frac {z^2}{C^3}-\frac {x^2}{A^3}\right)&V_{zx} +\frac
{xz}{AC}\left( \frac {V_{xx}}C-\frac {V_{zz}}A\right) +\frac
{zy}{BC^2} V_{xy}-\frac {yx}{BA^2}V_{yz}=0.
\endaligned
\end{equation}
This system is an analogue of the Bertrand-Darboux equation
(\ref{BerDar}) (see \cite{Drag2}).

\smallskip

Let
$$
\frac{x^2 C(A-C)}{z^2 (B-A)A}=\hat x, \quad
\frac{y^2C(C-B)}{z^2(B-A)B} = \hat y.
$$
The following statement was also proved in \cite{Drag2}:

\begin{theorem}\label{th:Vgama2}
For every $\gamma\in\mathbb C$, the function
$$
V_{\gamma}=(-\gamma +1)\left( \frac
{z^2}{x^2}\right)^{\gamma}F_4(1,-\gamma +2,2,-\gamma +1,\hat x,
\hat y)
$$
is a solution of the system {\rm (\ref{pde.sistem})}.
\end{theorem}

Thus, by solving Bertrand--Darboux equation and its
generalizations, as it is done in Theorems \ref{th:Vgama1} and
\ref{th:Vgama2}, one get large families of separable mechanical
systems with two degrees of freedom. It is well known that
separable systems of two degrees of freedom are necessarily of the
Liouville type, see \cite{Wh}.

\smallskip

Now the natural question of Poncelet type theorem describing
periodic solutions of such perturbed billiard systems arises. It
appears that again Darboux studied such a question, since in
\cite{Dar3}, he analyzed generalizations of Poncelet theorem in
the case of the Liouville surfaces.

\section*{Appendix 2\newline Poncelet Theorem on Liouville Surfaces}
\addtocontents{toc}{\newline}
\addtocontents{toc}{\numberline{}{\bf Appendix 2}\newline}
\addtocontents{toc}{\numberline{}{\bf Poncelet Theorem on
Liouville Surfaces\hfill43}\newline}

In this section, we are going to give the presentation and
comments to the Darboux results on the generalization of Poncelet
theorem to Liouville surfaces.

\subsection*{A2.1 Liouville Surfaces and Families of Geodesic
Conics}

In this subsection, following \cite{Dar3}, we are going to define
geodesic conics on an arbitrary surface, derive some important
properties of theirs and finally to obtain an important
characterization of Liouville surfaces via families of geodesic
conics.

\smallskip

Let $\mathcal C_1$ and $\mathcal C_2$ be two fixed curves on a
given surface $\mathcal S$. {\it Geodesic ellipses and hyperbolae}
on $\mathcal S$ are curves given by the equations:
$$
\displaylines{ \theta+\sigma=\const,\cr \theta-\sigma=\const,}
$$
where $\theta$, $\sigma$ are geodesic distances from $\mathcal
C_1$, $\mathcal C_2$ respectively.

\smallskip

A coordinate system composed of geodesic ellipses and hyperbolae
joined to two fixed curves is orthogonal. In the following
proposition, we are going to describe all orthogonal coordinate
systems with coordinate curves that can be regarded as a family of
geodesic ellipses and hyperbolae.

\begin{proposition} Let
\begin{equation}\label{ds}
ds^2=A^2du^2 + C^2dv^2
\end{equation}
be the surface element
corresponding to an orthogonal system of coordinate curves. Then
the coordinate curves represent a family of geodesic ellipses and
hyperbolae if and only if the coefficients $A$, $C$ satisfy a
relation of the form:
$$
\frac{U}{A^2} + \frac{V}{C^2} =1,
$$
with $U$ and $V$ being functions of $u$, $v$ respectively.
\end{proposition}

\begin{proof} By assumption, equations of coordinate
curves are:
$$
\theta+\sigma=\const, \quad \theta-\sigma=\const,
$$
with $\theta$, $\sigma$ representing geodesic distances from a
point of the surface to two fixed curves.

\smallskip

Thus:
$$
u=F(\theta+\sigma), \qquad v=F_1(\theta-\sigma).
$$
Solving these equations with respect to $\theta$ and $\sigma$, we
obtain:
$$
\theta=\phi(u)+\psi(v), \quad \sigma=\phi(u)-\psi(v).
$$

As geodesic distances, $\theta$ and $\sigma$ need to satisfy the
characteristic partial differential equation:
\begin{equation}\label{pde.geod}
\frac{A^2(\frac{\partial\xi}{\partial v})^2 +
C^2(\frac{\partial\xi}{\partial u})^2}{A^2C^2}=1.
\end{equation}
From there, we deduce the desired relation with $U=(\phi'(u))^2$,
$V=(\psi'(v))^2$.

\smallskip

The converse is proved in a similar manner.
\end{proof}

As a straightforward consequence, the following interesting
property is obtained:

\begin{corollary} If an orthogonal system of curves can
be regarded as a system of geodesic ellipses and hyperbolae in two
different ways, then it can be regarded as such a system in
infinitely many ways. \end{corollary}

Now, let us concentrate to Liouville surfaces, i.e.\ to surfaces
with the surface element of the form
\begin{equation}\label{element.liuvil}
ds^2=(U-V)(U_1^2du^2+V_1^2dv^2),
\end{equation}
where $U$, $U_1$ and $V$, $V_1$ depend only on $u$ and $v$
respectively.

\smallskip

Now, we are ready to present the characterization of Liouville
surfaces via geodesic conics.

\begin{theorem}\label{th:konike.liuvil}
An orthogonal system on a surface can be
regarded in two different manners as a system of geodesic conics
if and only if it is of the Liouville form. \end{theorem}

\begin{proof} Consider a surface with the element
(\ref{ds}). If coordinate lines $u=\const$, $v=\const$ can be
regarded as geodesic conics in two different manners then, by
Proposition 6, $A$, $C$ satisfy two different equations:
$$
\frac{U}{A^2} + \frac{V}{C^2} =1,\qquad \frac{U_1}{A^2} +
\frac{V_1}{C^2} =1.
$$
Solving them with respect to $A$, $C$, we obtain that the surface
element is of the form
$$
ds^2=\left(\frac{U}{U-U_1} + \frac{V}{V_1-V}\right)
\bigl((U-U_1)du^2+(V_1-V)dv^2\bigr).
$$

Conversely, consider the Liouville surface with the element
(\ref{element.liuvil}) and the following solutions of the equation
(\ref{pde.geod}):
$$
\displaylines{ \theta=\int U_1\sqrt{U-a}\,du + \int
V_1\sqrt{a-V}dv,\cr \sigma=\int U_1\sqrt{U-a}\,du - \int
V_1\sqrt{a-V}dv. }
$$
The equations $\theta+\sigma=\const$, $\theta-\sigma=\const$ will
define the coordinate curves. \end{proof}

\subsection*{A2.2 Generalization of Graves and Poncelet Theorems
to Liouville Surfaces}\label{liuvil}

We learned from Darboux \cite{Dar3} that Liouville surfaces are
exactly these having an orthogonal system of curves that can be
regarded in two or, equivalently, infinitely many different ways,
as geodesic conics. Now, we are going to present how to make a
choice, among these infinitely many presentations, of the most
convenient one, which will enable us to show the generalizations
of theorems of Graves and Poncelet. All these ideas of enlightening
beauty and profoundness belong to Darboux \cite{Dar3}.

\smallskip

Consider a curve $\gamma$ on a surface $\mathcal S$. The {\it
involute} of $\gamma$ with respect to a point $A\in\gamma$ is
the set of endpoints $M$ of all geodesic segments $TM$, such that:

$T\in\gamma$;

$TM$ is tangent to $\gamma$ at $T$;

the length of $TM$ is equal to the length of the segment
$TA\subset\gamma$; and

these two segments are placed at the same side of the point $T$.
\begin{figure}[h]
\centering
\begin{minipage}{0.6\textwidth}
\centering
\includegraphics[width=5cm,height=4cm]{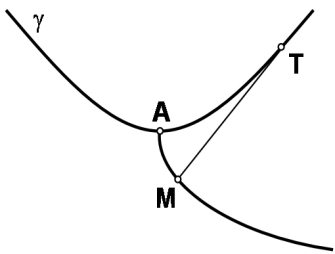}
\caption{Involute}
\end{minipage}
\end{figure}

\smallskip

Involutes have the following important property, which follows
immediately from the definition:

\begin{lemma}\label{lema:developande} The geodesic segments $TM$ are orthogonal to the
involute, and the involute itself is orthogonal to $\gamma$ at
$A$.
\end{lemma}

Now, we are going to find explicitly the equations of involutes
of coordinate curves on a Liouville surface $\mathcal S$ with the
surface element (\ref{element.liuvil}).

\begin{lemma}\label{lema:jne.devel}
The curves on $\mathcal S$ given by the equations:
\begin{equation}\label{jne.devel}
\aligned \theta&=\int U_1\sqrt{U-a}\,du + \int
V_1\sqrt{a-V}dv=\const,\cr \sigma&=\int U_1\sqrt{U-a}\,du - \int
V_1\sqrt{a-V}dv=\const
\endaligned
\end{equation} are involutes of the coordinate curve whose
parameter satisfies the equation:
$$
(U-a)(V-a)=0.
$$
\end{lemma}

\begin{proof} Fix the parameter $a$. The equations of of
geodesics normal to the curves $\theta=\const$, $\sigma=\const$
are obtained by differentiating (\ref{jne.devel}) with respect to
$a$:
\begin{equation}\label{geodezijske.normale}
\frac{U_1du}{\sqrt{U-a}}\pm\frac{V_1dv}{\sqrt{a-V}}=0.
\end{equation}
Let $u_0$ be a solution of the equation $U-a=0$. Then, the
geodesic line (\ref{geodezijske.normale}) will satisfy $du=0$ at
the point of intersection with the curve $u=u_0$, i.e.\ it will be
tangent to this coordinate curve. The statement now follows from
Lemma \ref{lema:developande}. \end{proof}

\begin{proposition} Coordinate curves on a Liouville surface are
geodesic conics with respect to any two involutes of one of
them.
\end{proposition}

\begin{proof} Follows from Lemma \ref{lema:jne.devel} and
the proof of Theorem \ref{th:konike.liuvil}.
\end{proof}

Now, we are ready to prove the generalization of Graves' theorem.

\begin{theorem}\label{th:graves}
Let $\mathcal E_0:u=u_0$ and $\mathcal E_1:u=u_1$ be coordinate
curves on the Liouville surface $\mathcal S$. For a point
$M\in\mathcal E_1$, denote by $MP$ and $MP'$ geodesic segments
that touch $\mathcal E_0$ at $Q$, $Q'$. Then the expression
$$
\ell(MP)+\ell(MP')-\ell(PP')
$$
is constant for all $M$, where $\ell(MP)$, $\ell(MP')$, and
$\ell(PP')$ denote lengths of geodesic segments $MP$, $MP'$, and
of the segment $PP'\subset\mathcal E_0$ respectively.
\begin{figure}[h]
\centering
\begin{minipage}[b]{0.6\textwidth}
\centering
\includegraphics[width=5cm,height=4cm]{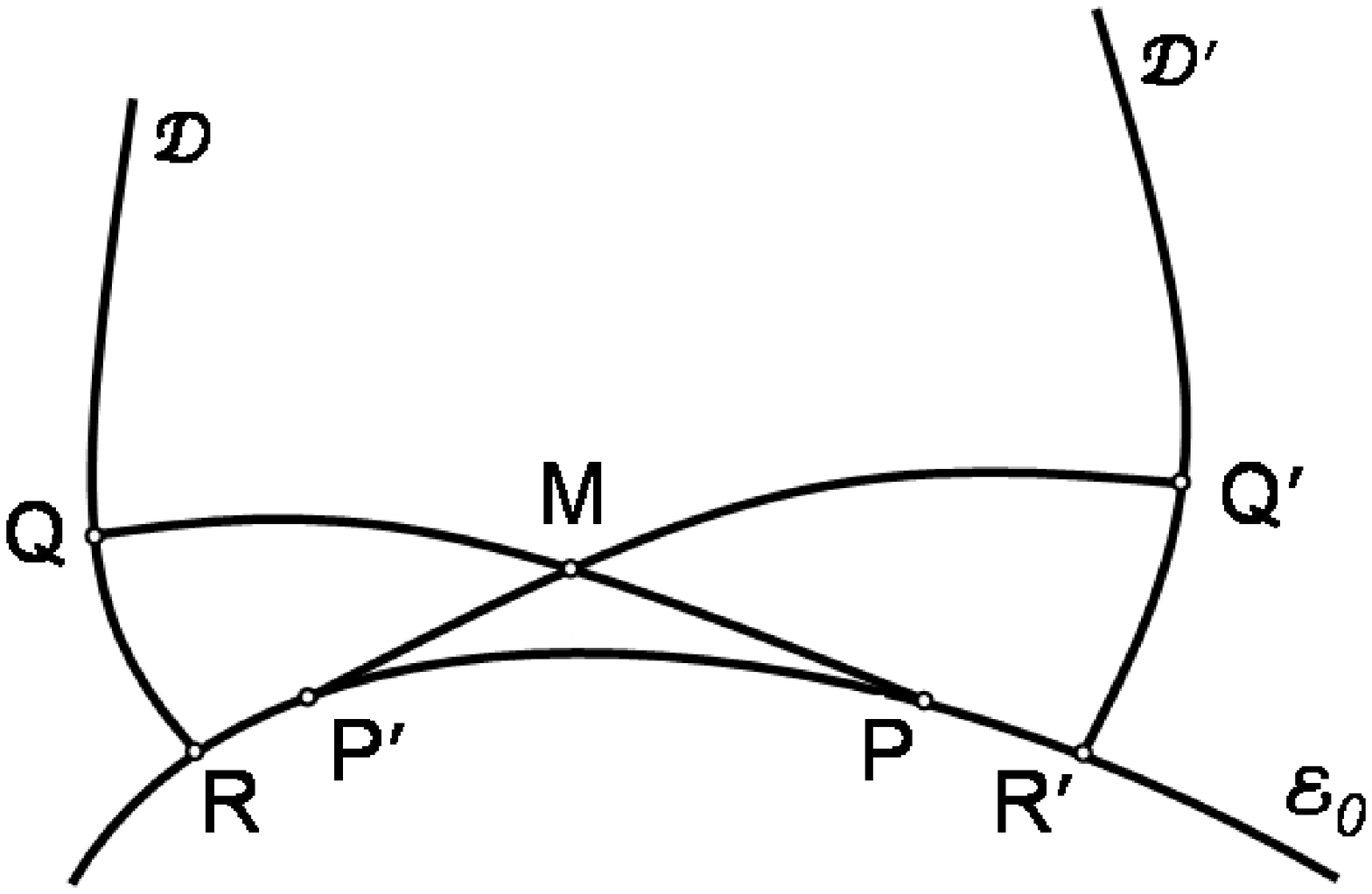}
\caption{Theorem \ref{th:graves}}
\end{minipage}
\end{figure}
\end{theorem}

\begin{proof} Let $\mathcal D$, $\mathcal D'$ be
involutes of the curve $\mathcal E_0$ with respect to points
$R,R'\in\mathcal E_0$, and $Q,Q'$ intersections of geodesics
$MP,MP'$ with these involutes.

\smallskip

Both $\mathcal E_0$ and $\mathcal E_1$ are geodesic ellipses with
base curves $\mathcal D$, $\mathcal D'$, thus the sum
$\ell(MQ)+\ell(MQ')$ remains constant when $M$ moves on $\mathcal
E_1$.

\smallskip

Since $\ell(PR)=\ell(MP)+\ell(MQ)$,
$\ell(P'R')=\ell(MP')+\ell(MQ')$, we have:
$$
\aligned \ell(MQ)+\ell(MQ')
&=\ell(PR)+\ell(P'R')-\ell(MP)-\ell(MP')\cr
&=\ell(RR')-\bigl(\ell(MP)+\ell(MP')-\ell(PP')\bigr),
\endaligned
$$
and the theorem is proved. \end{proof}

From here, the complete analogue of the Poncelet theorem can be
derived:

\begin{theorem} Let us consider a polygon on the Liouville
surface $\mathcal S$, with all sides being geodesics tangent to a
given coordinate curve, and each vertex but one moving on a
coordinate curve. Then the last vertex also remains on a fixed
coordinate curve. \end{theorem}

\subsection*{Acknowledgements}
\addtocontents{toc}{\numberline{}\newline}
\addtocontents{toc}{\numberline{}{\bf Acknowledgements
\hfill47}\newline\newline}

The research was partially supported by the Serbian Ministry of
Science and Technology, Project {\it Geometry and Topology of
Manifolds and Integrable Dynamical Systems}. The authors would
like to thank Prof.\ B.\ Dubrovin, Yu.\ Fedorov and S.\ Abenda for
interesting discussions, and to Prof.\ M.\ Berger for historical
remarks. One of the authors (M.R.) acknowledges her gratitude to
Prof.\ V.\ Rom-Kedar and the Weizmann Institute of Science for the
kind hospitality and support during in the final stage of the work
on this paper.

\addtocontents{toc}{\numberline{}{\bf References\hfill47}}

\end{document}